\documentclass[fleqn,usenatbib]{mnras}
\usepackage[english]{babel}
\usepackage{lmodern}
\usepackage{microtype}
\usepackage{booktabs}
\usepackage{bm}
\usepackage{graphicx}
\usepackage{amsfonts}
\usepackage{amsmath}
\usepackage{amssymb}
\usepackage{listings}
\usepackage{physics}
\usepackage{fancyhdr}

\newcommand{\DP}[1]{\mathrm{DP}\qty(#1)} 
\newcommand{\msun}{M_\odot} 

\numberwithin{equation}{section}

\title[(H)DPGMM]{(H)DPGMM: A Hierarchy of Dirichlet Process Gaussian Mixture Models for the inference of the black hole mass function}

\author[S. Rinaldi and W. Del Pozzo]{
Stefano Rinaldi$^{1,2}$\thanks{E-mail: stefano.rinaldi@phd.unipi.it}
and Walter Del Pozzo$^{1,2}$
\\
$^{1}$Dipartimento di Fisica ``E. Fermi", Università di Pisa, I-56127 Pisa, Italy\\
$^{2}$INFN, Sezione di Pisa, I-56127 Pisa, Italy
}
\date{\today}

\date{Accepted XXX. Received YYY; in original form ZZZ}

\pubyear{2021}

\begin{document}
\label{firstpage}
\pagerange{\pageref{firstpage}--\pageref{lastpage}}
\maketitle
\bibliographystyle{mnras}

\begin{abstract}
    We introduce (H)DPGMM, a hierarchical Bayesian non-parametric method based on the Dirichlet Process Gaussian Mixture Model, designed to infer data-driven population properties of astrophysical objects without being committal to any specific physical model. We investigate the efficacy of our model on simulated datasets and demonstrate its capability to reconstruct correctly a variety of population models without the need of fine-tuning of the algorithm. We apply our method to the problem of inferring the black hole mass function given a set of gravitational wave observations from LIGO and Virgo, and find that the (H)DPGMM infers a binary black hole mass function that is consistent with previous estimates without the requirement of a theoretically motivated parametric model. Although the number of systems observed is still too small for a robust inference, (H)DPGMM confirms the presence of at least two distinct modes in the observed merging black holes mass function, hence suggesting in a model-independent fashion the presence of at least two classes of binary black hole systems.
\end{abstract}


\section{Introduction}

Beginning with GW150914~\citep{150914discovery}, LIGO~\citep{LIGOdetector} and Virgo~\citep{VIRGOdetector} observed 50 compact binary coalescences: 11 during the first two observing run (O1, O2)\citep{GWTC1} and 39 during the first half of the third observing run (O3a)~\citep{GWTC2}. 

The detection of these coalescences opened a new window on the Universe: through gravitational waves (GWs), we now have the possibility to look directly into previously unobserved compact binaries and make measurements of the properties of these systems and of the nature of the objects that compose them.

In particular, focusing on the binary black hole systems (BBHs), we are able to measure the intrinsic parameters of the two black holes, masses and spins~\citep[e.g.][]{GWTC2}. The ensemble of inferred BBHs properties allow the characterisation of the merging black hole population, hence potentially shedding light onto the processes that lead to black hole binaries -- or, more broadly, black holes in general -- formation. The understanding of BBHs formation has profound implications for stellar evolution in general. For this reason, several efforts have been devoted to the determination of the formation channels for this kind of systems~\citep{abbott_150914, DiCarlo,kruckow,kumamoto,rodriguez:2015,rodriguez:2016}, considering both isolated evolution and dynamical capture scenarios. Given the different physical processes involved in the aforementioned scenarios, one expects two different BBHs mass and spin distributions, resulting in an observed one given by a mixture of the two, with weights set by the fraction of systems in the universe that have been formed accordingly. Another interesting possibility is that some of the detected systems contain at least one second-generation BH, hence resulting in a scenario in which at least some of the observed BBHs are the result of successive mergers~\citep{kimball, mapelli,hierarchicalmergers}.

Processes that happen during the evolution of the BHs stellar progenitors are believed to leave a peculiar signature in the black hole mass function: this is the case, for instance, for the pair-instability supernova (PISN), setting an upper limit to the progenitor star's core~\citep{bond, Heger_2003, pisn_on_mergers}, resulting in a cut-off on the BH mass function. The very same process, ultimately determined by the efficiency of the $^{12}$C$+\alpha\to \gamma + ^{16}$O nuclear reaction~\citep[][e.g.]{farmer:2020} and the efficiency of stellar winds mass loss~\citep{Belczynski:2020}, is expected to cause a pile-up at $\sim 40\ \msun$~\citep{pileup}. 

In addition to astrophysical processes and formation channels, there is the appealing possibility that some of the detected signals come from primordial black holes~\citep{franciolini:2021}, originated from the collapse of overdensities during the radiation-dominated early Universe~\citep{carrprimordial}.

Current observations seem to indicate that there is evidence for the fact that a single formation channel could not contribute for more than $70\%$ of the total observational samples; 
\citet{multiplechannels}, in fact, shows that single formation channel mass functions are disfavoured against more complex models that account for multiple channels.

From a more GW-centred perspective, the knowledge of the black hole mass function can be exploited, for instance, in setting a more informative prior for parameter estimation~\citep{Veitch:2015} or for cosmological analyses~\citep{O2cosmo}.
The inference of the Hubble constant using the so-called statistical method~\citep{schutz,delpozzo:2012,Abbott:2019yzh,gray:2020} requires, while handling selection effects, the knowledge of the black hole population properties. The impact of wrong population assumptions on cosmological measurements are discussed in~\citet{cosmodark}.
In addition to this, some papers suggests that some (eventual) peculiar features of the mass function, especially in the high-mass region, can lead to cosmological constraints: this is the case, for example, of the PISN cut-off~\citep{h0at8}.

Population properties of black holes and neutron stars (NSs) detected by LIGO and Virgo are discussed in~\citet{pop1} and, more recently, in~\citet{pop2}. These two papers deal with the inference of the black hole mass distribution using parametric population models. In particular,~\citet{pop2} makes use of four different population models, with different degrees of complexity. Features in these models were introduced to accommodate different expected black hole formation channels.

Given the increasing number of GW events being detected, the concrete possibility of new unforeseen classes of systems being among the observed set of merging binaries, it is imperative to be able to infer the population properties without being committal towards specific model-dependent prescriptions. Several efforts towards this direction exist~\citep{Li, ay, vamana, binnedmandel}, but, to the best of the authors' knowledge, no full non-parametric scheme has yet been proposed. 

Non-parametric Bayesian methods are powerful tools to perform inference without the need to specify a model~\citep[see][part V]{gelman}. The reader should not be confused by the nomenclature \emph{non-parametric} since Bayesian non-parametrics does not imply that the underlying model has no parameter, but that the number of parameters is, in fact, countably infinite. The existence and well-posedness of such schemes heavily rely on the De Finetti theorem for exchangeable sequences~\citep{definetti}.
A pure non-parametric method has the interesting property of being able to reconstruct the probability density just letting the data speak for themselves: the retrieved distribution is the one that best describes the observations.
This flexibility, however, comes at the cost that no direct information on the underlying physics is provided by these non-parametric approaches, e.g. the inferred distribution is entirely phenomenological.
In order to explain the physics behind any kind of observed feature in the recovered distribution, one still has to rely on parametric models built on formation channels.

This paper presents a semi-hierarchical non-parametric technique based on the Dirichlet Process Gaussian Mixture Model~\citep{escobar&west, rasmussen}. Our scheme is constructed to infer the population properties of merging binaries without the need to specify a population model. The method we propose differs from classical Hierarchical Dirichlet Process methods~\citep[e.g.][]{tehetal} 
since we relax the interdependence between categories assumed in the latter. This allows for a more efficient exploration of the posterior distribution allowing for a massively parallel first inference step.

The paper is organised as follows: in Sec.~\ref{sec:(H)DPGMM} we present the statistical framework and we introduce the Hierarchy of Dirichlet Process Gaussian Mixture Models along with an outline of how to implement the inference algorithm using Collapsed Gibbs sampling, presented in Sec.~\ref{sec:collapsed}. In Sec.~\ref{sec:simulations} we test our model on three different simulated situations, gradually approaching one of the models presented in~\cite{pop2}. Sec.~\ref{sec:undersampling} discusses some limitations of our non-parametric approach, while Sec.~\ref{sec:seleff} introduces a way to handle selection effects. Finally, in Sec.~\ref{sec:gwtc} we apply our model to data from GWTC--2.

\section{A Hierarchy of Dirichlet Process Gaussian Mixture Models}\label{sec:(H)DPGMM}

Before introducing the Dirichlet Process and our hierarchical generalisation, following~\citet{skyloc}, we first briefly review the Dirichlet distribution.
\subsection{Dirichlet Distribution}
Consider an experiment whose outcome can assume $k$ known different values, like rolling a die or tossing a coin. We are interested in inferring the probability distribution for these $k$ categories using $N$ samples. Denote with $n_i$ the number of times we observed category $i$, while $q_i$ is the probability associated with the same category. Knowing the probabilities $\mathbf{q} \equiv \{q_1,\dots, q_k\}$, the probability of observing $n_1,\ldots,n_k$ is given by a multinomial distribution:
\begin{equation}\label{multinomial}
    p(n_1,\dots, n_k|q_1,\dots, q_k) = \frac{N!}{n_1!\dots n_k!}\prod_{i = 1}^kq_i^{n_i}\,.
\end{equation}
Using Bayes' theorem, we can infer the probabilities $\mathbf{q}$ given an array of observations $\mathbf{n} \equiv \{n_1,\dots, n_k\}$ as
\begin{equation}\label{posterior}
    p(\mathbf{q}|\mathbf{n}) = \frac{p(\mathbf{n}|\mathbf{q})p(\mathbf{q})}{\int p(\mathbf{n}|\mathbf{q})p(\mathbf{q})\dd\mathbf{q}}
\end{equation}
where $p(\mathbf{q})$ is the prior probability over the probability assignments. 
Being conjugate to the multinomial distribution, a common prior choice for this problem is the Dirichlet distribution:
\begin{equation}\label{dirichletdist}
    p(\mathbf{q}|\boldsymbol\alpha) = \mathrm{Dir}(\mathbf{q}|\boldsymbol\alpha) =\frac{\Gamma(A)}{\prod_{i=1}^k\Gamma(\alpha_i)}\prod_{i=1}^k q_i^{\alpha_i-1}\,,
\end{equation}
where we defined $\boldsymbol\alpha \equiv \{\alpha_1,\dots, \alpha_k\}$ and $A = \sum_{i=1}^k \alpha_i$, with the only requirement that $\alpha_i > 0\ \forall i$. The set of $\alpha_i$ are called \emph{concentration parameters} and control the shape of the distribution, hence the relative prior belief for the number of outcomes in each of the categories. The concentration parameters effectively quantify how we believe the probabilities $\mathbf{q}$ will be assigned.

In order to clarify this, consider a die game, where the number of possible outcomes is 6. Assuming that we have no reason to believe that the die is unfair, we wish to assign equal a priori probability to each outcome. In this case, the concentration parameters will have the same value, $\alpha_1 = \ldots = \alpha_6 = \alpha$. This choice leads to the so-called \emph{symmetric Dirichlet distribution}. A crucial role is played by the magnitude of the quantity $\alpha$: larger values lead to distributions which are concentrated around the uniform distribution $q_1 = \ldots = q_6 = 1/6$, while $\alpha \gtrsim 1$ implies that we still believe that every side has the same probability of being rolled but we admit for the possibility that our die could be unfair. 

In fact, $\alpha = 1$ corresponds to the uniform probability distribution on the $\mathbf{q}$ simplex, meaning that every possible probability assignment is \emph{a priori} equally likely: we are expressing complete ignorance on the state of the system, hence we are using a non-informative prior~\citep{gelman}. On the other hand, $\alpha < 1$ still gives a symmetric distribution but its shape tells a completely different story: we are sure that our die is unfair \emph{but we don't know which side is favoured}.

We now want to explore the scenario in which we know that the number 6 corresponds to the biased side of the die. In this case, we would like to encode this knowledge in the prior distribution, hence stepping back from the symmetric Dirichlet distribution. For simplicity, we assume that only one side is favoured while the others are equally likely. In this case, the concentration parameter associated with the number 6 will be larger than the others. The ratio $\alpha_6/\alpha_{i\neq 6}$ expresses the disparity between $q_6$ and $q_{i\neq 6}$. Also in this case, the magnitude of these coefficients expresses our belief in this probability assignment.

The role of the quantity $A = \sum_{i=1}^k\alpha_i$ is to determine how concentrated (or sparse) the distribution over distributions is around the specified prior choice for $\mathbf{q}$.

The Dirichlet distribution has the convenient property of being the conjugate prior  to the multinomial likelihood~\eqref{multinomial}, meaning that the posterior distribution will be a Dirichlet distribution as well. This property simplifies both computation and the inclusion of new data in the analysis~\citep{gelman}. With this prior choice, Eq.~\eqref{posterior} becomes
\begin{equation}\label{explicitposterior}
    p(\mathbf{q}|\mathbf{n},\boldsymbol\alpha) = \frac{1}{\mathcal{N}}\prod_{i=1}^k q_i^{(n_i+\alpha_i) -1} \propto \mathrm{Dir}(\mathbf{q}|\mathbf{n}+\boldsymbol\alpha)\,.
\end{equation}
from which we can compute the expected value of probability $q_i$ as
\begin{equation}
    \mathbb{E}[q_i] = \frac{a_i + n_i}{A+N}\,.
\end{equation}
In the limit of $n_i\gg a_i$ this result approaches the intuitively expected value $n_i/N$.
\subsection{Dirichlet Process}
When dealing with rolling of dice, political election polls, etc., the number of categories can be arbitrarily large, but it is always finite and known in advance. Can we somehow generalise the Dirichlet distribution to allow for the possibility that the number of outcomes is unknown and therefore (potentially) countably infinite? 

Such generalisation is possible and was introduced in~\citet{ferguson} with the introduction of a stochastic process in the space of probability distributions: the Dirichlet Process (DP).
Consider an experiment whose outcome is a real number in an interval $I$ and a probability density function $G$ over $I$, $G$ is said to be DP distributed if for any partition of $I$ $I_1,\ldots, I_n$\footnote{E.g., $n$ different histogram bins. The only requirement for the partition is, again, to have an arbitrarily large number of subintervals.}, $\mathbf{G} = (G(I_1),\ldots, G(I_n))$ is distributed according to a (finite) Dirichlet distribution.
Introducing a base distribution $H$ over $I$ and a concentration parameter $a$, $\mathbf{H} = (H(I_i),\ldots, H(I_n))$ takes the role of the previously defined $\boldsymbol\alpha$ in determining which probability distribution we expect to see \emph{a priori}, while $a$ determines the concentration of the Dirichlet distribution.
\begin{equation}
    \mathbf{G} \sim \mathrm{Dir}(a\mathbf{H})\,
\end{equation}
and, taking the limit of (countably) infinite subintervals, the probability density $G$ is distributed according to a DP:
\begin{equation}
    G \sim \DP{a,H}\,.
\end{equation}
For a more formal definition of the DP, we refer the reader to~\citet{Teh}. 
The base distribution $H$ can be interpreted as the mean, or expected value, of the DP: if one takes several realisations of the stochastic process,
\begin{equation}
    \mathbb{E}[G(I_i)] = H(I_i)\,.
\end{equation}
Note that, however, it is incorrect to say that $\mathbb{E}[G] = H$, even in the limit $a\to\infty$\footnote{This limit for the concentration parameter states that we are sure that our prior choice is the right one.}: this is due to the fact that the distribution $G$ is discrete while $H$ can be (and usually is) smooth. 
The DP has the same conjugacy properties of the Dirichlet distribution, hence we can use it as prior distribution like in \eqref{explicitposterior}. Assume that we have $N$ different draws $x_i$ from $G$. Given these draws and the fact that $G\sim\DP{a,H}$, the posterior distribution for $G$ is~\citep{Teh}:
\begin{equation}
    G|x_1,\ldots, x_N \sim \DP{\alpha + N, \frac{\alpha}{\alpha+N}H + \frac{N}{\alpha+N}\frac{\sum_{i=1}^N\delta_{x_i}}{N}}\,.
\end{equation}
Here, $\delta_{x_i}$ denotes the point mass located at $x_i$ and $\sum_{i=1}^N\delta_{x_i}$ is often called the \emph{empirical distribution}. This expression for the posterior distribution highlights also the role of the concentration parameter: it can be interpreted as the weight of the prior strength expressed in terms of number of draws from the underlying distribution $G$. If $N\gg\alpha$, the empirical distribution becomes a good approximation of $G$ and the inference becomes independent from the prior.~\citet{Teh} also gives an expression for the predictive distribution of a new draw $x_{N+1}$,
\begin{equation}
    x_{N+1}|x_1,\ldots, x_N \sim \frac{1}{\alpha+N}\qty(\alpha H+\sum_{i=1}^N\delta_{x_i})\,,
\end{equation}
which will be useful in what follows.

\subsection{Dirichlet Process Gaussian Mixture Model}\label{sec-DPGMM}
We have seen that a realisation of a DP is a \emph{discrete} probability distribution, hence it is atomic. Since we are interested in inferring smooth probability densities, we will make use of a Dirichlet Distribution Gaussian Mixture Model (DPGMM)~\citep{escobar&west}. The idea is to use the DP as prior probability for the parameters of a smoothing kernel function family.
The probability distribution for $x$ can be written as
\begin{equation}\label{DPGMM}
   p(x) = \sum_{k=1}^\infty w_k N(x|\mu_k,\sigma^2_k)G(w_k,\mu_k,\sigma^2_k)\,,
\end{equation}
where $N$ denotes the normal density function, $\mu$ and $\sigma^2$ are its mean and variance\footnote{For simplicity of notation we concentrate on a one dimensional kernel. The generalisation to multivariate normal distributions is straightforward.}  and $w$ is the weight or \emph{mixing proportion} associated with the mixture component $k$. In particular, $\sum_{k=1}^\infty w_k = 1$.

There are several ways to construct samples from a DP~\citep{Teh}, depending on the property of the DP one wants to emphasise. Here we follow~\citet{skyloc} in using the so-called \emph{stick-breaking} construction~\citep{sethuraman} where
\begin{equation}\label{weights}
    w_i = \beta_i\prod_{j=1}^{i-1}(1-\beta_j)
\end{equation}
and
\begin{equation}\label{beta}
    \beta_i \sim \mathrm{Beta}(1,\alpha)\,,
\end{equation}
where $\alpha$ is the concentration parameter of the DP. The Beta distribution is
\begin{equation}
    \mathrm{Beta}(\beta|a,b) = \frac{\Gamma(a+b)}{\Gamma(a)\Gamma(b)}\beta^{a-1}(1-\beta)^{b-1}\,.
\end{equation}
The combination of~\eqref{weights} and~\eqref{beta} leads to the Griffiths--Engen--McCloskey (GEM) distribution~\citep{pitman}, hence we can write
\begin{equation}
    \mathbf{w} \sim \mathrm{GEM}(\alpha)\,,
\end{equation}
where $\textbf{w}$ denotes all the weights $w_i$.

A common and convenient choice as prior for $\mu$ and $\sigma^2$ is the Normal--Inverse Gamma (NIG) distribution~\citep{gorur}
\begin{equation}
    p(\mu,\sigma^2|a, b, m, V) = N(\mu|m,V\sigma^2)\Gamma^{-1}(\sigma^2|a,b)\,,
\end{equation}
where the Inverse Gamma distribution is
\begin{equation}
    \Gamma^{-1}(\zeta|a,b) = \frac{b^a}{\Gamma(a)}\frac{e^{-\frac{b}{\zeta}}}{\zeta^{a+1}}\,.
\end{equation}
The NIG is conjugate to the Gaussian, and its use as prior on $\mu$ and $\sigma^2$ allows us to marginalise out analytically these parameters. Different priors can be used, e.g.~\citet{gorur}. In order to keep things as general as possible, in the following of this section we will not specify any particular distribution for the base distribution, denoting it $F(\mu,\sigma^2|\Theta)$, where $\Theta$ is any set of parameters required by the prior distribution.

We need also to specify a prior distribution for the concentration parameter $\alpha$. Following~\citet{gorur}, we choose the Gamma distribution
\begin{equation}
    \mathrm{Gamma(\zeta|\beta, \gamma)} = \frac{\gamma^\beta}{\Gamma(\beta)}\zeta^{\beta-1}e^{-\gamma \zeta},
\end{equation}
in particular $\alpha \sim \mathrm{Gamma}(\alpha^{-1}|1,1)$.

In summary, the prior DPGMM model can be expressed as:
\begin{align}
    &\alpha \sim \mathrm{Gamma(\alpha^{-1}|1,1)},\\
    &\mathbf{w} \sim \mathrm{GEM}(\alpha),\\
    &\mu_i,\sigma^2_i \sim F(\mu,\sigma^2|\Theta),\label{priorsmusigma}\\
    &x \sim \sum_{i = 1}^\infty w_i N(x|\mu_i,\sigma^2_i).
\end{align}

Let us imagine having a set of observations $\mathbf{x} = \{x_1,\ldots,x_N\}$ drawn from $G$. According to \eqref{DPGMM}, each of them will be drawn from one of the infinite mixture components $k$, selected with probability $w_k$. Let us consider for a moment a simpler case in which the number of components is finite, say $K$. In the following, $\mathbf{w} \equiv \{w_1,\ldots, w_K\}$. From~\eqref{multinomial}, the probability for the occupation number $n_k$ is multinomial.

It is useful to introduce indicator variables $\mathbf{z} \equiv \{z_1,\ldots z_N\}$ that denote to which mixture component each of the $x_i$ belongs to. The distribution of these variables is~\citep{gorur}
\begin{equation}
    p(\mathbf{z}| \mathbf{w}) = \prod_{i=1}^K w_i^{n_i}\,.
\end{equation}
Using a symmetric Dirichlet prior $\alpha_i = \alpha/K$ we can integrate out the mixing proportions $\mathbf{w}$ and obtain a probability distribution for the indicator variables:
\begin{equation}
    p(\mathbf{z}|\alpha) = \frac{\Gamma(\alpha)}{\Gamma(N+\alpha)}\prod_{i=1}^K \frac{\Gamma(n_i+\alpha/K)}{\Gamma(\alpha/K)}\,.
\end{equation}
This probability distribution allows us to write the conditional probability for a new data point $x_{N+1}$ being drawn from component $j$:
\begin{equation}
    p(z_{N+1} = j| \mathbf{z},\alpha) = \frac{p(z_{N+1}=j,\mathbf{z}|\alpha)}{p(\mathbf{z}|\alpha)} = \frac{n_j + \alpha/K}{N+\alpha}\,.
\end{equation}
Now, taking the limit $K \to \infty$, the conditional probability for $z_{N+1}$ becomes
\begin{equation}
    p(z_{N+1} = j| \mathbf{z},\alpha) = \frac{n_j}{N+\alpha}
\end{equation}
for an already populated component and
\begin{equation}
    p(z_{N+1} \neq j\ \forall j \in \mathbf{z}|\mathbf{z},\alpha) = \frac{\alpha}{N+\alpha}
\end{equation}
for all other empty components combined.

Up to now, we derived the probability for a data point to be associated with a component conditioned only to the occupation number of each component. Consider a different situation: we have the same data set $\mathbf{x}$ as before, and we know from which mixture component each data point comes. Let us add a new data point $x_{N+1}$ to our set without the information about the mixture component it has been drawn from; we want to derive the conditional probability for the new point to be associated with mixture component $j$:
including the parameters of the prior on $\mu_j$ and $\sigma^2_j$ (denoted as $\Theta$) we get
\begin{multline}\label{conditional_x}
    p(z_{N+1} = j | x_{N+1}, \mathbf{x},\mathbf{z}, \alpha, \Theta) =\\= \frac{p(x_{N+1}|\mathbf{x},\mathbf{z}, z_{N+1} = j, \alpha, \Theta)p(z_{N+1}=j|\mathbf{z},\alpha)}{\mathcal{N}},
\end{multline}
where

\begin{multline}\label{clustering_x}
    p(x_{N+1}|\mathbf{x},\mathbf{z}, z_{N+1} = j, \alpha, \Theta) =\\= \frac{p(x_{N+1},\mathbf{x}|\mathbf{z}, z_{N+1} = j, \alpha, \Theta)}{p(\mathbf{x}|\mathbf{z}, z_{N+1} = j, \alpha, \Theta)} =\\= \frac{\int p(x_{N+1},\mathbf{x}|\mu_j,\sigma^2_j,\mathbf{z}, z_{N+1} = j, \alpha, \Theta)p(\mu_j,\sigma_j|\Theta)\dd\mu_j\dd\sigma^2_j}{\int p(\mathbf{x}|\mu_j,\sigma^2_j,\mathbf{z}, z_{N+1} = j, \alpha, \Theta)p(\mu_j,\sigma_j|\Theta)\dd\mu_j\dd\sigma^2_j}.
\end{multline}
Since we are considering Gaussian kernels,~\eqref{clustering_x} becomes
\begin{multline}\label{p_x}
    p(x_{N+1}|\mathbf{x},\mathbf{z}, z_{N+1} = j, \alpha, \Theta) =\\=
    \frac{\int N(x_{N+1}|\mu_j,\sigma_j)\prod_{i|z_i=j} N(x_i|\mu_j,\sigma^2_j)F(\mu_j,\sigma^2_j|\Theta)\dd\mu_j\dd\sigma^2_j}{\int\prod_{i|z_i=j} N(x_i|\mu_j,\sigma^2_j)F(\mu_j,\sigma^2_j|\Theta)\dd\mu_j\dd\sigma^2_j}\,,
\end{multline}

where $i|z_i = j$ denotes all the data points associated with component $j$.

\subsection{A Hierarchy of DPGMMs}\label{sec-HDPGMM}
In the previous section we described the case in which our data points were drawn from the underlying distribution. Making a step further, imagine that we do not have direct access to $x_i$ but only to a set of samples $\mathbf{y_i} \equiv \{y_1,\ldots,y_m\}$ around $x_i$, drawn from a probability distribution $p(x_i)$. In this picture, we have no direct information about $\mathbf{x}$, our data set being $\mathbf{Y} \equiv \{\mathbf{y}_1,\ldots,\mathbf{y}_N\}$\footnote{In order to put a bit of context in this: every black hole in our Universe has a mass (the DP realisation $x_i$) and with LIGO and Virgo we are able, for each gravitational wave event, to draw mass samples from their posterior distribution around this mass value (the vector $\mathbf{y}_i$).}. In general, $p(x_i)$ could be any probability distribution: we are interested here in the case in which it can be modelled as a realisation of a DP.

According to~\citet{tehetal}, a Hierarchical Dirichlet Process (HDP) can be specified as:
\begin{align}
    &G_0\ |\ \gamma,H \sim\ \DP{\gamma,H},\\
    &G\ |\ \alpha, G_0 \sim\ \DP{\alpha, G_0},\\
    &\mathbf{Y}\sim\ G.
\end{align}
$\mathbf{Y}$ is a realisation of a single Dirichlet Process, where $\mathbf{x}$ corresponds to $G_0$. However, in this paper we deal with a slightly different problem: each $\mathbf{y}_i$ is a realisation of a different DP, in which $x_i$ enters as a parameter of the base distribution.
We can specify this process as 
\begin{align}
    &G' \sim\ \DP{\gamma, H},\\
    &x_i \sim\ G',\\
    &G_i\ |\ \alpha_i, G_0, x_i \sim\ \DP{\alpha_i, G_0(x_i)},\\
    &\mathbf{y}_i \sim\ G_i.
\end{align}
Due to this different structure, we refer to this process as a hierarchy of DPs, rather than a HDP.
Since we are, once again, interested in inferring smooth probability densities, we will make use, in this hierarchical picture, of a smoothing Gaussian kernel: hence the name \emph{Hierarchy of Dirichlet Process Gaussian Mixture Models}, or (H)DPGMM. We will refer to the $\mathbf{x}$-generating process as the \emph{outer} DPGMM while the $\mathbf{y}_i$-generating processes will be the \emph{inner} DPGMMs.

Given $\mathbf{Y}$, Eq.~\eqref{conditional_x} becomes

\begin{multline}\label{hierarchical}
    p(z_{N+1} = j | \mathbf{z}, \alpha,\Theta, \mathbf{y}_{N+1}, \mathbf{Y}) =\\= \int p(z_{N+1} = j |x_{N+1}, \mathbf{x}, \mathbf{z}, \alpha,\Theta) \prod_{i=1}^{N+1}p(x_i| \boldsymbol\theta_i)p(\boldsymbol\theta_i| \mathbf{y}_i)\dd x_i\dd\boldsymbol\theta_i,
\end{multline}
where $\boldsymbol\theta_i$ denotes all the parameters required by the probability distribution. In the specific case in which we assume $p(x_i)$ to be a realisation of a DPGMM,~\eqref{hierarchical} reads
\begin{multline}\label{HDPGMM}
    p(z_{N+1} = j | \mathbf{z}, \alpha,\Theta, \mathbf{y}_{N+1}, \mathbf{Y}) =\\= \int p(z_{N+1} = j |x_{N+1}, \mathbf{x}, \mathbf{z}, \alpha,\Theta) \prod_{i=1}^{N+1}\sum_{k=1}^\infty w_{k,i}N(x_i|\mu_{k,i},\sigma^2_{k,i})\times\\\times G_i(w_{k,i},\mu_{k,i},\sigma_{k,i}^2|\mathbf{y}_i)\dd x_i\dd w_{k,i}\dd \mu_{k,i}\dd \sigma^2_{k,i}.
\end{multline}
\section{Inference using Collapsed Gibbs sampling}\label{sec:collapsed}
The aim of this paper is to provide a method to infer the black hole mass function using gravitational wave observations. The rationale for our models comes from the following considerations: black hole masses in our Universe can be thought as a realisation of a DP whose base distribution is the mass function, while single-event GW mass samples are samples from the probability distribution $p(x_i)$. In the language of the previous section, black hole masses are the vector $\mathbf{x}$, the mass function is the base distribution $H$ and GW samples corresponds to $\mathbf{y}_i$.

There are several way to explore DPGMMs: a possibility is to use Gibbs sampling~\citep{neal, gorur}, another is the variational algorithm applied in~\citet{skyloc}. Our method relies on the former.

Imagine having a set of variables $\xi_1,\ldots,\xi_n$ that follow a multivariate probability distribution $p(\xi_1,\ldots,\xi_n)$ we want to draw samples from. The Gibbs sampling is a Markov chain Monte Carlo (MCMC) algorithm that finds its application when the joint probability distribution $p(\xi_1,\ldots,\xi_n)$ is difficult, computationally expensive or even impossible to evaluate but, at the same time, conditional probabilities 
\begin{equation}\label{conditional}
    p(\xi_i|\boldsymbol\xi_{-i})= p(\xi_i|\xi_1,\ldots, \xi_{i-1},\xi_{i+1},\ldots,\xi_n)
\end{equation}
are relatively easy to compute. Following~\citet{gorur}, $\boldsymbol\xi_{-i}$ denotes the vector $\boldsymbol\xi\equiv \{\xi_1,\ldots,\xi_n\}$ without the $i-$th element $\xi_i$.

In a nutshell, Gibbs sampling works as follows: beginning from a certain\footnote{It could be a randomly selected state as well as some pre-determined state: the initial state does not affect the algorithm outcome, as long as the chain is long enough.} initial state $\bar\xi_1,\ldots,\bar\xi_n$, it iteratively draws a new $\xi_i$ value from the conditional~\eqref{conditional}, keeping every other $\xi_j$ fixed, and updates its value. 

This algorithm could require the introduction of auxiliary variables. Considering the case of assigning data to different components of a Gaussian Mixture Model, indicator variables $\mathbf{z}$ and data point values $\mathbf{x}$ alone are not enough to evaluate the probability of a certain data point to be assigned to a specific component: we have to introduce the auxiliary variables that correspond to mean, variance and weights of the different components of the Gaussian Mixture Model and include them in our sampling routine\footnote{See Eqs.~\eqref{conditional_x} and~\eqref{clustering_x}: with the Gibbs sampling, the integral over $\mu_j$ and $\sigma^2_j$ is evaluated via Monte Carlo integration.}.

We use a modification of the Gibbs sampling called Collapsed Gibbs Sampling~\citep{liu} in which some of the variables can be marginalised out in the conditional distribution. This is the case, for example, in which we specify a conjugate prior: $F(\mu,\sigma^2|\Theta) = \mathrm{NIG}(\mu,\sigma|\Theta)$~\citep{gorur}. Here $\Theta = \{m, V, a, b \}$ is a placeholder for the parameters of the NIG prior.

Taking into account the situation described in Sec.~\ref{sec-DPGMM}, we can draw a sample from a DP given a set of data points $\mathbf{x}$ and their indicator variables $\mathbf{z}$ specifying the conditional probability distribution for $\mathbf{w}$, $\boldsymbol\mu = \{\mu_1,\ldots,\mu_K\}$ and $\boldsymbol\sigma^2 = \{\sigma^2_1, \ldots, \sigma^2_K \}$, where $K$ is the number of components with $n_i \geq 1$. From~\eqref{explicitposterior} with the prescription of a symmetric Dirichlet distribution, 
\begin{equation}\label{p_w}
    \mathbf{w}\ |\ \mathbf{z}, \alpha \sim \frac{1}{\mathcal{N}} \prod_{i=1}^{K}w_i^{(n_i + \alpha/K)-1},
\end{equation}
while, making use of the conjugate prior,
\begin{equation}\label{p_musigma}
    \mu_j,\sigma_j\ |\ \mathbf{z},\mathbf{x},\Theta \sim \mathrm{NIG}(\mu_j,\sigma_j|\{x_i|z_i = j\},\Theta)
\end{equation}
for each $j = 1,\ldots,K$.

The update of $z_i$ is made selecting each element of $\mathbf{z}$ one at a time and dealing with it as it is a brand new data point which needs to be assigned to a component of our mixture.
Naming $p^i_j = p(z_i = j|x_i,\mathbf{z}_{-i},\mathbf{x}_{-i}, \alpha,\Theta)$ the quantity in Eq.~\eqref{hierarchical}, 
\begin{equation}\label{p_z}
    z_i\ |\ \mathbf{z}_{-i} \sim Discrete(p_1^i,\ldots,p_K^i,p_{new}^i),
\end{equation}
where $p_{new}^i$ denotes the probability for the data point $i$ to be assigned to an empty component. Thanks to the fact that we chose the conjugate prior, the integral in Eq.~\eqref{p_x} is analytically treatable and becomes a Student--t distribution~\citep{murphy}.

Last thing we need to update is the concentration parameter $\alpha$.~\citet{gorur} provides the conditional likelihood for $\alpha$:
\begin{equation}
p(\alpha|K) = \frac{\Gamma(\alpha)}{\Gamma(N+\alpha)}\alpha^K \,.
\end{equation}
With the inclusion of the prior, we get
\begin{equation}\label{p_alpha}
    \alpha\ |\ \mathbf{z} \sim \frac{\Gamma(\alpha)}{\Gamma(N+\alpha)}\alpha^K e^{-\frac{1}{\alpha}} \,.
\end{equation}

As a general summary, some pseudo-code should look something like this:
\begin{lstlisting}[label=verb1, caption = DPGMM,frame=tb, mathescape = true,escapeinside={(*}{*)}]
initialise $\mathbf{z}$, $\alpha$
iterate:
    for $z_i$ in $\mathbf{z}$:
        draw $z_i$ from (*\eqref{p_z}*)
        update $z_i$
    draw $\mathbf{w}$ from (*\eqref{p_w}*)
    for j from 1 to K:
        draw $\mu_j$, $\sigma^2_j$ from (*\eqref{p_musigma}*)
    save $\mathbf{w}$, $\boldsymbol\mu$, $\boldsymbol\sigma^2$
    draw $\alpha$ from (*\eqref{p_alpha}*)
    update $\alpha$
\end{lstlisting}
In principle, the collapsed Gibbs sampling scheme can be applied also to the situation described in~\ref{sec-HDPGMM}. The main difference between the algorithm we just presented and a scheme for the (H)DPGMM is that the latter accounts for $p(x_i)$, so the parameters of the $N$ DPGMMs we use to describe these probability distributions must be included in our sampling scheme. For the sake of brevity, we will denote all of these parameters with $\boldsymbol\Lambda \equiv \{\Lambda_1,\ldots,\Lambda_N\}$, where $\Lambda_i \equiv \{\mathbf{w}, \boldsymbol\mu, \boldsymbol\sigma^2\}$ for a single DPGMM.

More than this, due to the fact that we are marginalising over $x_i$ with a realisation from a DPGMM as probability density, choosing a NIG prior on $\mu$ and $\sigma^2$ does not help since it is not conjugate to the likelihood anymore. We use an uniform prior in $[\mu_{min}, \mu_{max}]$ and $[\sigma^2_{min},\sigma^2_{max}]$. Eq.~\eqref{p_musigma} becomes
\begin{equation}\label{p_musigma_hierarchical}
    \mu_j,\sigma_j\ |\ \mathbf{z},\boldsymbol\Lambda \sim \frac{1}{\mathcal{N}}\prod_{i=1}^{n_j}\sum_{k=1}^{K_i} w_{k,i} N(\mu_{k,i}|\mu_j,\sigma^2_j).
\end{equation}
With the notable exception that now $p_j^i$ is given by Eq.~\eqref{HDPGMM}, the conditional probability distribution for the indicator variables $\mathbf{z}$ is the same we derived before, as well as the one for the concentration parameter of the outer DPGMM $\gamma$.

We report here some pseudo-code for the inference of an (H)DPGMM:
\begin{lstlisting}[label=verb2, caption = (H)DPGMM,frame=tb, mathescape = true,escapeinside={(*}{*)}]
initialise $\mathbf{z}$, $\gamma$
iterate:
    for $\Lambda_i$ in $\boldsymbol\Lambda$:
        update $\Lambda_i$ following Listing (*\ref{verb1}*)
    for $z_i$ in $\mathbf{z}$:
        draw $z_i$ from (*\eqref{p_z}*)
        update $z_i$
    draw $\mathbf{w}$ from (*\eqref{p_w}*)
    for j from 1 to K:
        draw $\mu_j$, $\sigma^2_j$ from (*\eqref{p_musigma_hierarchical}*)
    save $\mathbf{w}$, $\boldsymbol\mu$, $\boldsymbol\sigma^2$
    draw $\gamma$ from (*\eqref{p_alpha}*)
    update $\gamma$
\end{lstlisting}
We implemented an algorithm, based on this pseudo-code, to explore the (H)DPGMM. The Python code is available at \url{https://github.com/sterinaldi/hdp-population}.

\subsection{Pre-processing}
The statistical method and the algorithm we presented above can be used to reconstruct any kind of probability density. However, given the fact that we are approximating the distribution with a sum of Gaussian distributions, better results are achieved with Gaussian-like probability distributions.

In order to obtain a smoother and easier-to-approximate probability distribution we follow~\citet{golomb, talbot}, where the authors suggest a coordinate change to map the distribution into a better behaved domain. 

Given a uniform probability distribution for the outer variable $x$ between $x_{min}$ and $x_{max}$, the posterior samples $y_i$ are mapped to the $[0,1]$ interval using the cumulative density function (CDF) of the prior distribution $F(x)$. For a uniform prior we have
\begin{equation}
   F(x) = \frac{x - x_{min}}{x_{max}-x_{min}}.
\end{equation}
These values can be interpreted as quantiles for a Gaussian distribution centred in 0 with $\sigma^2 = 1$. The sample $\eta_i$, which is the image of the sample $y_i$ in the new space, reads
\begin{equation}
    \eta_i = \Phi^{-1}\qty(F(y_i)),
\end{equation}
where $\Phi^{-1}$ is the probit function, the inverse CDF of the normal distribution.

One can use the samples $\boldsymbol\eta \equiv\{\eta_1,\ldots,\eta_N\}$ in this new space to approximate the probability distribution $\mathcal{G}(\chi)$, where $\chi = \Phi^{-1}\qty(F(x))$: since we are interested in approximating $\mathcal{F}(x)$, 
\begin{equation}
\mathcal{F}(x) = \frac{\mathcal{G}(\chi)}{N(\chi| 0,1)}.
\end{equation}
This coordinate change, however, is ill-defined for $x = x_{min}, x_{max}$. The prior boundaries must therefore be selected with care in order to avoid issues in proximity of these values.

\section{Simulations}\label{sec:simulations}
In order to demonstrate the effectiveness of the (H)DPGMM at inferring the black hole mass function, we applied our model to three different sets of simulated GW observations. We did not undertake the exercise of a end-to-end simulation campaign, but we generate mock posterior distributions for each simulated GW event.
In this section we are not dealing with any issue related with selection effects\footnote{Selection effects are all these biases that ensure that a particular set of samples is not representative of the true underlying distribution. Talking about gravitational waves, this reflects the fact that our detectors have different sensivities in different regions of the population's parameter space.}. Here we assume that every event is detectable and detected, postponing the discussion about how to account for selection effects in the subsequent section.

Prior and initial state choices for the two levels of (H)DPGMM are:
\begin{enumerate}
    \item Inner DPGMM:
    \begin{enumerate}
        \item $\mathbf{z}$ initial state: the sorted mass samples are sliced in $K_0$ different components. We set $K_0 = 5$;
        \item Prior on $\mu$ and $\sigma^2$~\eqref{priorsmusigma}:  NIG with parameters $a = 1$, $V = 1$, $m = \sum_i t_i/N$ the mean of the samples and $b = a(s_0^2/16)$ with $s_0 = (\sum_i(t_i-m)^2/(N-1))^{1/2}$.
        We put an upper bound on the $\sigma^2$ prior at $\sigma_{max}^2 = s_0^2/4$. 
    \end{enumerate}
    \item Outer DPGMM:
    \begin{enumerate}
        \item $\mathbf{z}$ initial state: the events are sorted according to the mean of the associated set of samples and then sliced in $K_0$ different components. We set $K_0 = 5$;
        \item Prior on $\mu$ and $\sigma^2$~\eqref{priorsmusigma}: uniform on the rectangle $[t_{min}, t_{max}]\times[\sigma^2_{min},\sigma^2_{max}]$, with $\sigma^2_{min} = (s_0/16)^2$, $\sigma_{max}^2 = (s_0/3)^2$ and $t_{min}, t_{max}$ equals to the minimum and maximum, respectively, of all of the single event samples, while $s_0$ is the standard deviation of these samples.
    \end{enumerate}
\end{enumerate}
All the specified parameters are intended in the transformed space, not in physical space.

Setting an upper bound on $\sigma^2$ for the inner DPGMM in principle prevents us from using the conjugate prior properties. However, here we are assuming that the contribution of the region with $\sigma^2 > \sigma_{max}^2$ to the integral in Eq.~\eqref{p_x} is negligible, hence we approximately recover the conjugacy properties.

This choice is motivated empirically; while investigating simulated populations of BBHs we observed that if the variance of the inner DPGMM is allowed to grow unconstrained, among the solutions explored by the algorithm we found that sometimes all data points would be associated with a single Gaussian component with small concentration parameter, hence giving a poor fit to the observed posterior samples histograms. The reason for this behaviour can be understood in terms of the clustering properties of the DPGMM: since the DPGMM looks for the distribution which maximises the predictive likelihood~\eqref{clustering_x} with the smallest number of components, whenever possible it will try to put the greatest number of samples in a component as large as possible, in order to keep all the samples nearby the mean of the distribution\footnote{Here, \emph{nearby} loosely means \emph{within one or two standard deviations}.}. 

We want to emphasise the fact that our method does not require the user to predetermine and set the number of Gaussian components that enters the mixture.
The use of a Dirichlet Process as prior allows us to account for a countably infinite number of components, even if only a finite number of them have samples associated and hence are represented: thus, the degree of freedom associated to the number of active component is not fixed.
The value $K_0$ we set is an arbitrary choice for the \emph{initial} state. New components are added to the mixture every time a sample is assigned to a new cluster or removed once the last sample is associated to a different cluster: hence, as long as the chain is long enough, the inferred distribution is not affected by this choice.
We tested the robustness of this method with different $K_0$ and we found that the results are stable against the variation of the initial number of components.

As a measure of how much the reconstructed probability distribution differs from the simulated one, we use the Jensen--Shannon distance (see the introduction in~\citet{JensDist})
\begin{equation}
    \mathrm{JSD}(p||q) = \sqrt{\frac{D_{KL}(p||m)+D_{KL}(q||m)}{2}},
\end{equation}
where $m(x) = \frac{p(x)+q(x)}{2}$ and $D_{KL}$ is the Kullback--Leibler divergence~\citep{KLdiv}, defined as 
\begin{equation}
    D_{KL}(f||g) = \int_{-\infty}^{\infty} f(x)\log(\frac{f(x)}{g(x)})\dd x.
\end{equation}
\subsection{DPGMM}
\begin{figure}
    \centering
    \includegraphics[width=\columnwidth]{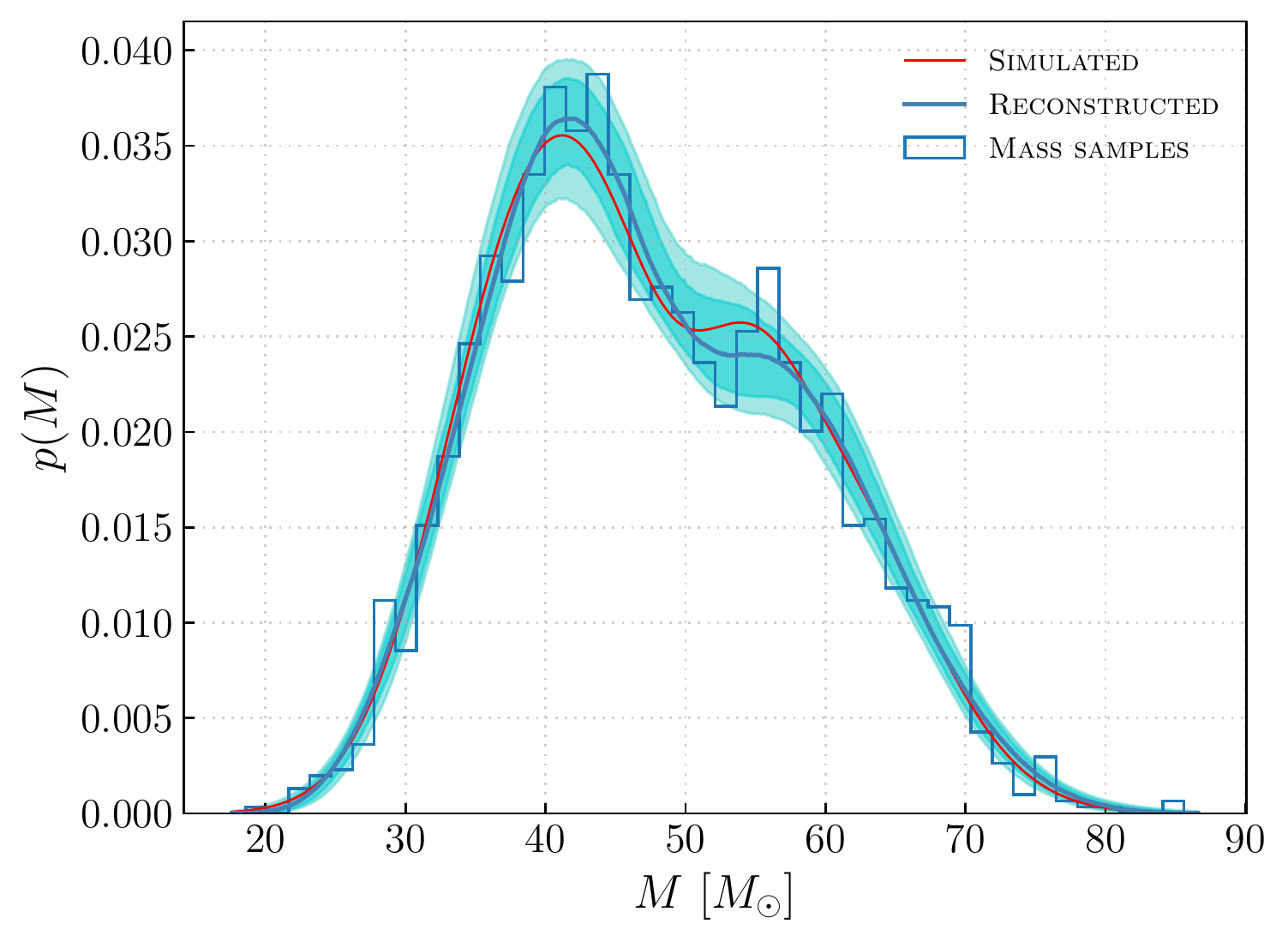}
    \caption{Reconstructed and simulated probability densities. The histogram shows the mass samples.}
    \label{fig:mockpost}
\end{figure}

As a first check, we want to demonstrate that the inner DPGMM properly reconstructs probability densities functions from sample. For this purpose, we simulated a mock posterior distribution, inspired by a ``typical'' event mass posterior from GWTC--2 as a weighted sum of Gaussians (see Table~\ref{tab:mockpost}). From this distribution we draw 2000 mass samples.
\begin{table}
    \caption{Simulated probability density parameters.}
    \centering
    \begin{tabular}{c|ccc}
    \toprule
        Component&$w_i$&$\mu_i\ [M_\odot]$& $\sigma_i\ [M_\odot]$\\
    \midrule
        1& 0.4 & 38 & 6\\
        2& 0.1 & 54 & 4\\
        3& 0.2 & 45 & 5\\
        4& 0.3 & 60 & 7\\
    \bottomrule
    \end{tabular}
    \label{tab:mockpost}
\end{table}
We used these samples to draw 1000 realisations of the posterior function using our Collapsed Gibbs sampling. The reconstructed probability distribution is shown in Figure~\ref{fig:mockpost}.
We find that the reconstructed probability density is very close to the simulated distribution, with the Jensen--Shannon distance of $\mathrm{JSD} = 3.4^{+1.0}_{-1.3}\cdot 10^{-2}$ nats.
\subsection{Bimodal Gaussian}\label{sec:bimodal}
We now turn to the investigation of our (H)DPGMM by analysing a toy model in which the mass function is a mixture of two Gaussian distributions
\begin{equation}\label{eq:bimodal}
    p(M) = \frac{N(M|\mu_1, \sigma^2_1)+N(M|\mu_2, \sigma^2_2)}{2},
\end{equation}
with $\mu_1 = 25\ \msun$, $\mu_2 = 55\ \msun$, $\sigma_1^2 = 16\ \msun^2$ and $\sigma^2_2 = 25\ \msun^2$. 
We then sample 200 masses from \eqref{eq:bimodal} and draw mass samples for event $i$ from a Gaussian distribution with mean $M_i \sim p(M)$ and standard deviation $\sigma_i$ from a flat-in-log distribution between $3\ \msun$ and $5\ \msun$. 
We use this procedure for every simulation presented in this paper.

\begin{figure}
    \centering
    \includegraphics[width = \columnwidth]{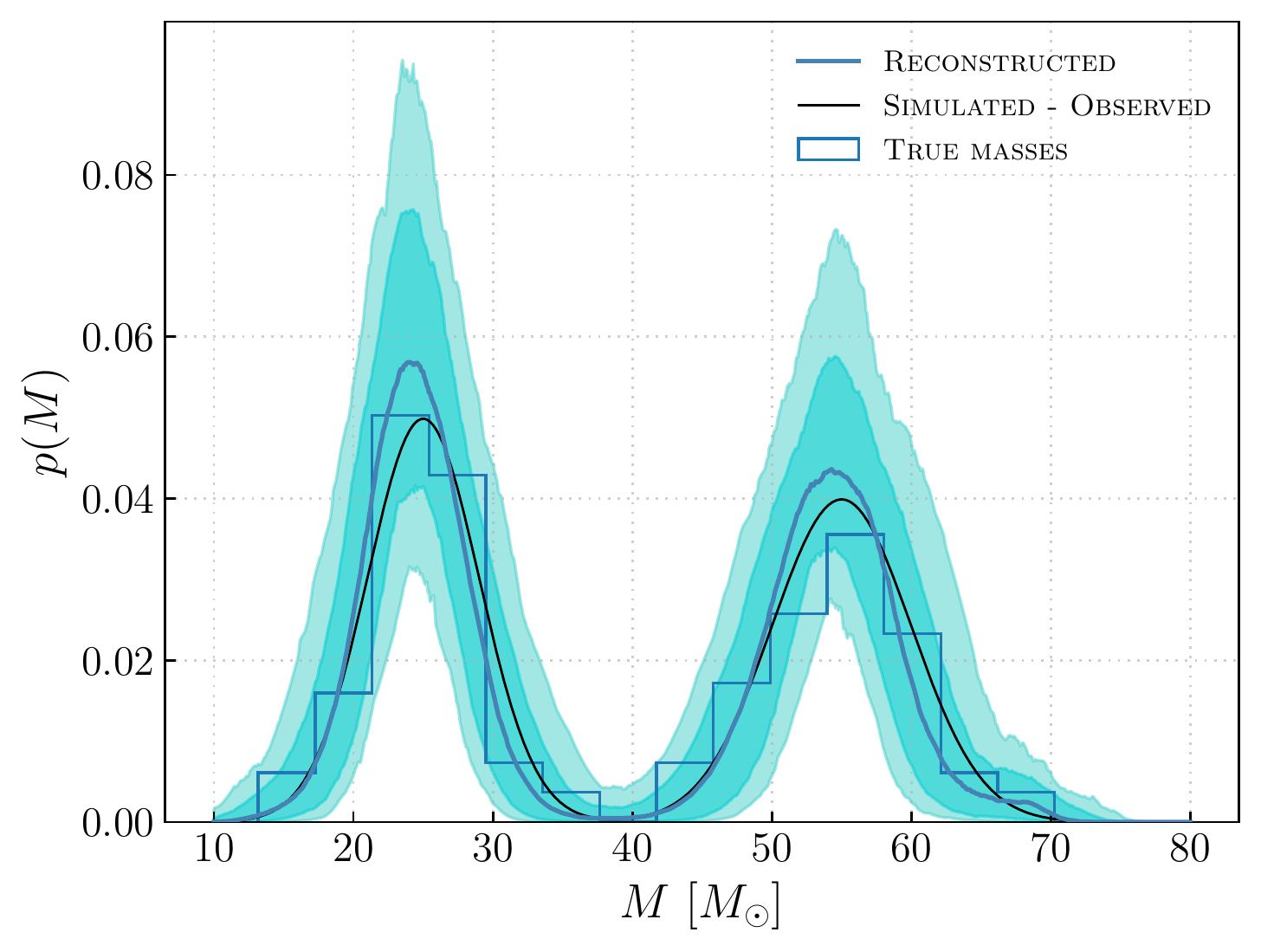}
    \caption{Reconstructed and simulated probability density for the bimodal Gaussian case.}
    \label{fig:bimodal}
\end{figure}

The reconstructed probability density is shown in Figure~\ref{fig:bimodal} and it is compatible with~\eqref{eq:bimodal}.
In this case, $\mathrm{JSD} = 0.19^{+0.12}_{-0.09}\ \mathrm{nats}$. 
\subsection{Power law}\label{sec:pl}
\begin{figure}
    \centering
    \includegraphics[width = \columnwidth]{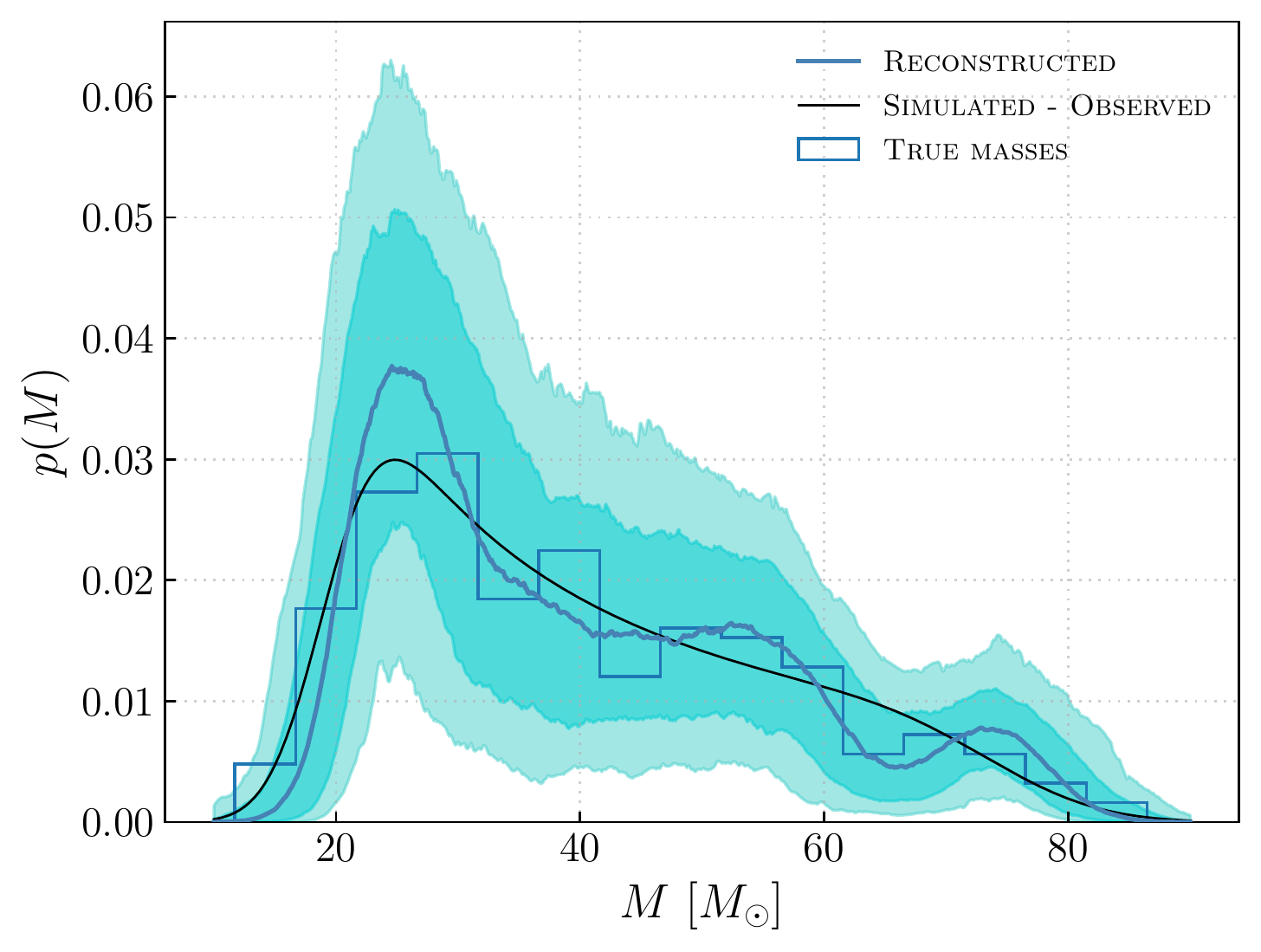}
    \caption{Reconstructed and simulated probability density for the tapered power law case.}
    \label{fig:PL}
\end{figure}

Having established that (H)DPGMM behaves correctly on simple mixture of Gaussians, we turn our attention on the more complicated -- and realistic -- case of a mass function which is a tapered power law
\begin{multline}\label{taperedPL}
    p(M) = PL(M) =\\= M^{-\alpha}\frac{\qty(1+\erf\qty(\frac{M-M_{min}}{\lambda_{min}}))\qty(1+\erf\qty(\frac{M-M_{max}}{\lambda_{max}}))}{4},
\end{multline}
with $\alpha = 1.2$, $M_{min} = 20\ \msun$, $M_{max} = 75\ \msun$, $\lambda_{min} = 5\ \msun$, $\lambda_{max} = 10\ \msun$. Here,
\begin{equation}
    \erf(x) = \frac{2}{\sqrt{\pi}}\int_0^x e^{-t^2}\dd t.
\end{equation}
Our catalogue is composed of 250 events drawn from this distribution. 
Here we have more events than the previous simulation: this choice comes from the fact that the underlying distribution is somewhat more complex than a bimodal Gaussian. 200 events were not enough to grasp all the features of the mass function. 
We will discuss the effects a limited data set in Sec.~\ref{sec:undersampling}.
The reconstructed mass function is shown in Figure~\ref{fig:PL}.
Unsurprisingly, the mass function reconstruction is less accurate than the previous case. However, the simulated probability density is found within the $90\%$ credible interval. The Jensen--Shannon distance is $\mathrm{JSD} = 0.21^{+0.11}_{-0.08}\ \mathrm{nats}$.
\subsection{Power law + Gaussian peak}
\begin{figure}
    \centering
    \includegraphics[width = \columnwidth]{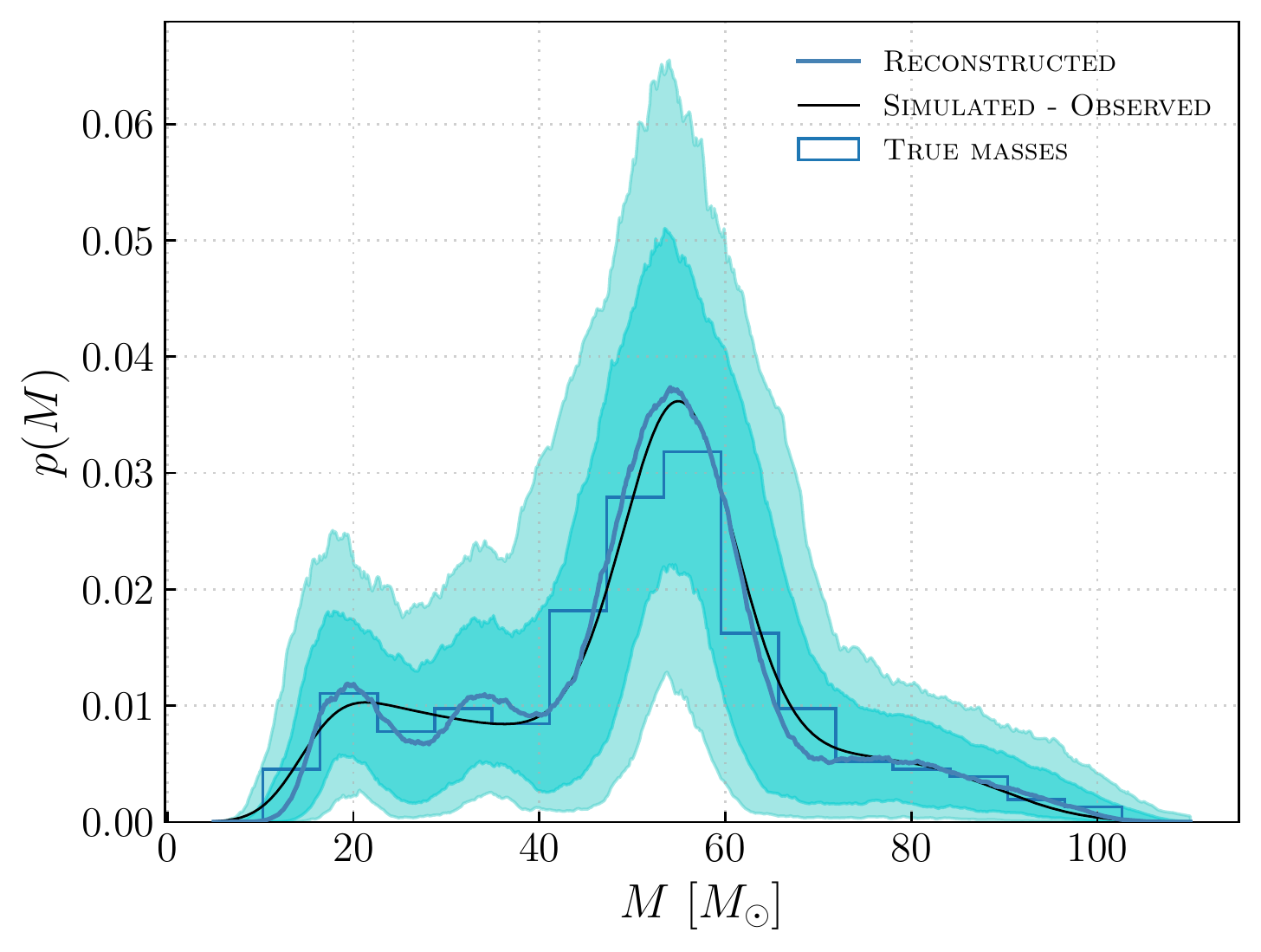}
    \caption{Reconstructed and simulated probability density for the \textsc{Power law + Peak} case.}
    \label{fig:plpeak}
\end{figure}

As last test, we simulated a mass function similar to the \textsc{Power law + Peak} from~\citet{pop2} as
\begin{equation}\label{eq:plpeak}
    p(M) = bN(M|\mu,\sigma^2) + (1-b)PL(M).
\end{equation}
Table~\ref{tab:plpeak} reports the parameters for this probability density.
\begin{table}
    \caption{Parameters for probability density~\eqref{eq:plpeak}}
    \centering
    \begin{tabular}{cccc}
    \toprule
        $b$&$\mu\ [\msun]$ &$\sigma^2\ [\msun^2]$ & $\alpha$  \\
    \hline
        0.9&55 &36 &0.5  \\
    \toprule
        $M_{min}\ [\msun]$&$\lambda_{min}\ [\msun]$ &$M_{max}\ [\msun]$ &$\lambda_{max}\ [\msun]$  \\
        \hline
        15&5 &90 &10\\
    \bottomrule
    \end{tabular}
    \label{tab:plpeak}
\end{table}
Here as well our simulated catalogue contains 250 events. The reconstructed probability density is displayed in Figure~\ref{fig:plpeak}.
The Jensen--Shannon distance, in this case, is $\mathrm{JSD} = 0.22^{+0.14}_{-0.08}\ \mathrm{nats}$.

\section{Interpreting the (H)DPGMM distribution}
Assuming a parametric model for our inference implies including a certain amount of information, which comes from the previous knowledge of the physical phenomenon we are interested in. On the other hand, non-parametric models like (H)DPGMM are data-driven models, meaning that all the available information comes from the data themselves. This approach leads to a very flexible model, whose interpretation requires the assumption that the data are indeed representative of the underlying process. Since (H)DPGMM will reconstruct the distribution that best accommodates the observed data, if this assumption fails we cannot interpret the recovered probability density as a good estimate of the \emph{real} mass function.
We will discuss in what follows two examples: (i) the case in which one has few data points available; (ii) the case in which selection effects are present.

\subsection{Undersampling}\label{sec:undersampling}

\begin{figure}
    \centering
    \includegraphics[width = \columnwidth]{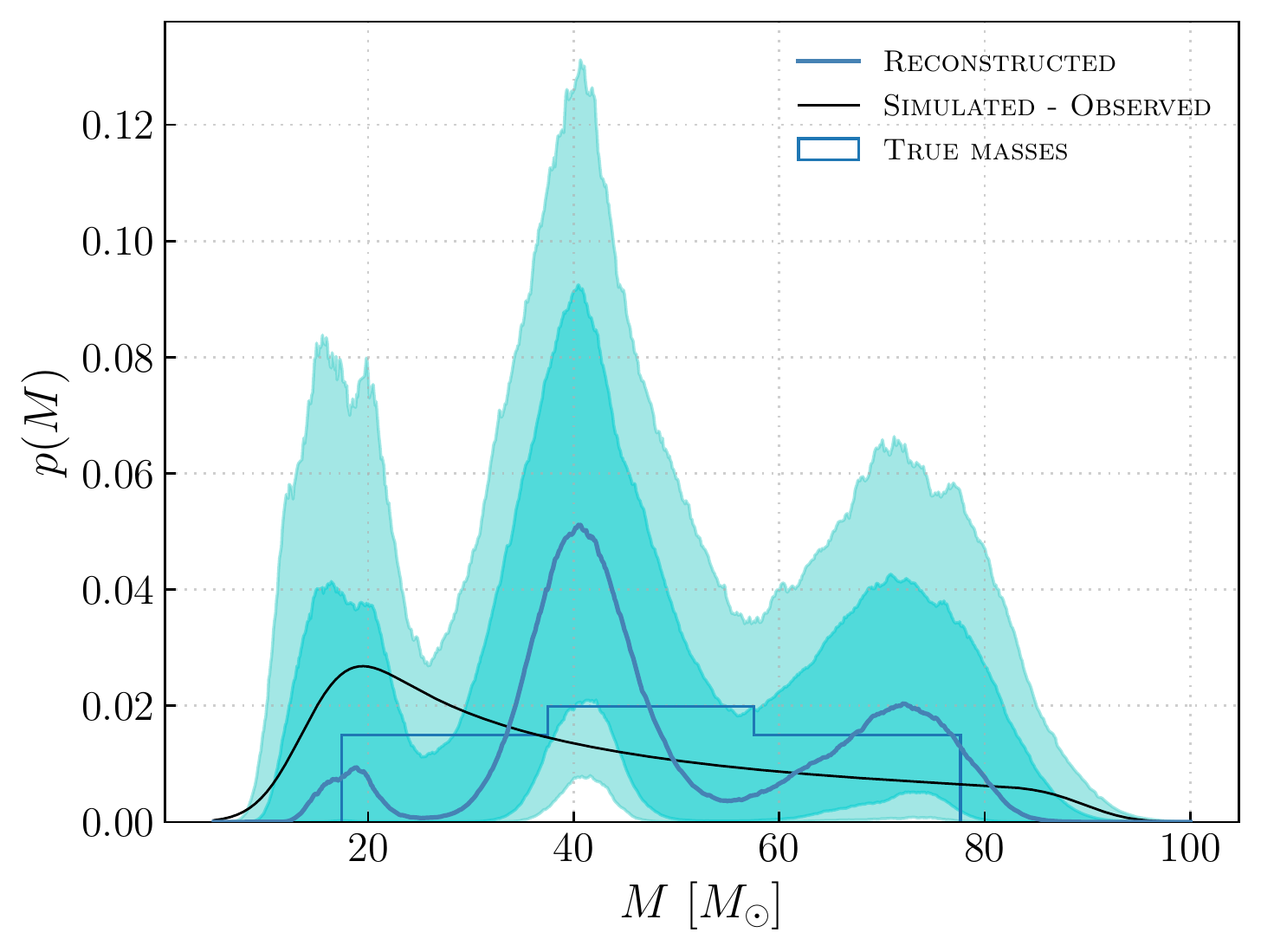}
    \caption{Simulated and reconstructed probability distribution for the tapered power law case with a small number of events.}
    \label{fig:wrong}
\end{figure}

This simulation uses the same probability distribution as~\ref{sec:pl}, Eq.~\ref{taperedPL}, with $\alpha = 1.1$, $M_{min} = 15\ \msun$, $M_{max} = 90\ \msun$, $\lambda_{min}=5\ \msun$, $\lambda_{max} = 10\ \msun$. However, in this case we considered for our analysis only 10 events. The results of this simulation are shown in Figure~\ref{fig:wrong}.
We see that the reconstructed probability distribution does not qualitatively agree with the simulation; although the reconstructed credible regions are very broad and encompass the underlying mass function, the (H)DPGMM reconstructed distribution shows clear signs of multimodality, potentially leading the naive reader to believe in the presence of 3 distinct populations. For this case, we find $\mathrm{JSD} = 0.45^{+0.11}_{-0.11}\ \mathrm{nats}$.
This simple experiment suggests that drawing physically meaningful conclusions from non-parametric models such as (H)DPGMM needs to be done with care whenever the data set under analysis is limited in size.

\subsection{Incorporating selection effects}\label{sec:seleff}
In all simulations presented in Section~\ref{sec:simulations}, we deliberately neglected selection effects, hence we assumed that all events are observable and observed. In this subsection we turn our attention to their effect on the (H)DPGMM inference. Since each of the mass functions we reconstructed in Section~\ref{sec:simulations} is based exclusively on observed data, we will refer to it as the \emph{observed} mass function. On the other hand, following~\citet{pop2}, we will call \emph{astrophysical} mass function the ‘‘real'' distribution. Using Bayes' theorem and denoting with $O$ the fact that the events have been observed,
\begin{equation}\label{seleff}
    p(M) = \frac{p(M|O)p(O)}{p(O|M)},
\end{equation}
where $p(M|O)$ is the observed mass function and $p(O|M)$ is the probability of observing an event with mass $M$. As customary, we refer to $p(O|M)$ as the selection function $S(M)$. A detailed discussion of the selection function can be found in~\citet{selectionfunction}.

Since the selection function appears at the denominator,~\eqref{seleff} is valid only where $p(O|M)\neq 0$. This is reasonable we do not expect to be able to gather information about regions of the parameter space where the observing probability is zero.

In the context of gravitational wave astronomy, the selection function depends, at leading order, on the signal-to-noise ratio (SNR) $\rho$ which, in turn, depends on the mass of the black hole as well as on several other parameters $\theta$. Marginalising out all the nuisance parameters,
\begin{equation}
    S(M) =\int\Theta(\rho > \rho_{th})p(\rho|\theta,M)p(\theta)\dd \theta \dd \rho,
\end{equation}
where $\rho_{th}$ is the threshold SNR above which a signal is detected. For a more detailed discussion about how to derive a selection function, see~\citet{farrseleff}.
Here we are neglecting the possibility of having uncertainties on the selection function itself that, in principle, should be included in the analysis.

This simulation aims to demonstrate that it is possible, given the knowledge of the selection function and a set of observations, to reconstruct the mass function. In this case we use a bimodal Gaussian distribution (see~\eqref{eq:bimodal}) with $\mu_1 = 25$, $\mu_2 = 55$, $\sigma_1 = 4$ and $\sigma_2 = 5$  as astrophysical mass distribution.

\begin{figure}
    \centering
    \includegraphics[width = \columnwidth]{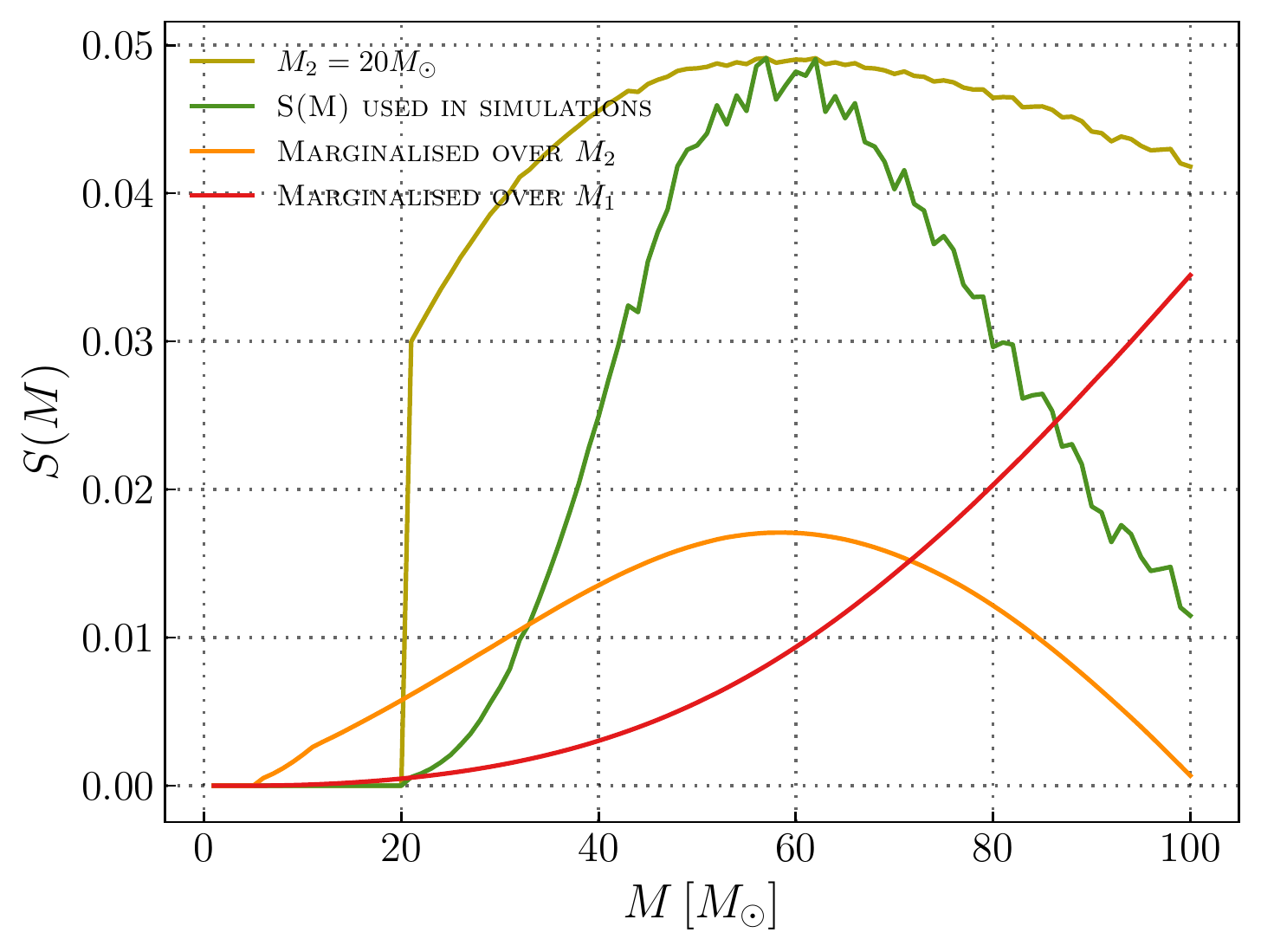}
    \caption{Selection function used in this paper and from~\citet{selectionfunction}.}
    \label{fig:selfunc}
\end{figure}

\citet{selectionfunction} provides the detection probability as a function of the primary mass when the secondary mass is fixed. Here we use their selection function $S_{V}$ assuming that all the events have $M_2 = 20\ \msun$. The corresponding selection function is reported in Figure~\ref{fig:selfunc}.

We see that $S_V$ is a slowly varying function of $M_1$. In order to enhance the effect of the selection function, we use
\begin{equation}
    S(M) = (S_V(M))^9\frac{\max\qty[{S_V(M)}]}{\max\qty[{(S_V(M))^9}]}
\end{equation}
In the previous simulations each event was directly drawn from the mass function. This time, we account for selection effects using the accept--reject sampling method: given a mass $\tilde M$ drawn from the astrophysical distribution $p(M)$, we accept the event with a probability $S(M)$. We generated $O(10^4)$ mass values as to obtain a total of 200 observed events.
\begin{figure}
    \centering
    \includegraphics[width = \columnwidth]{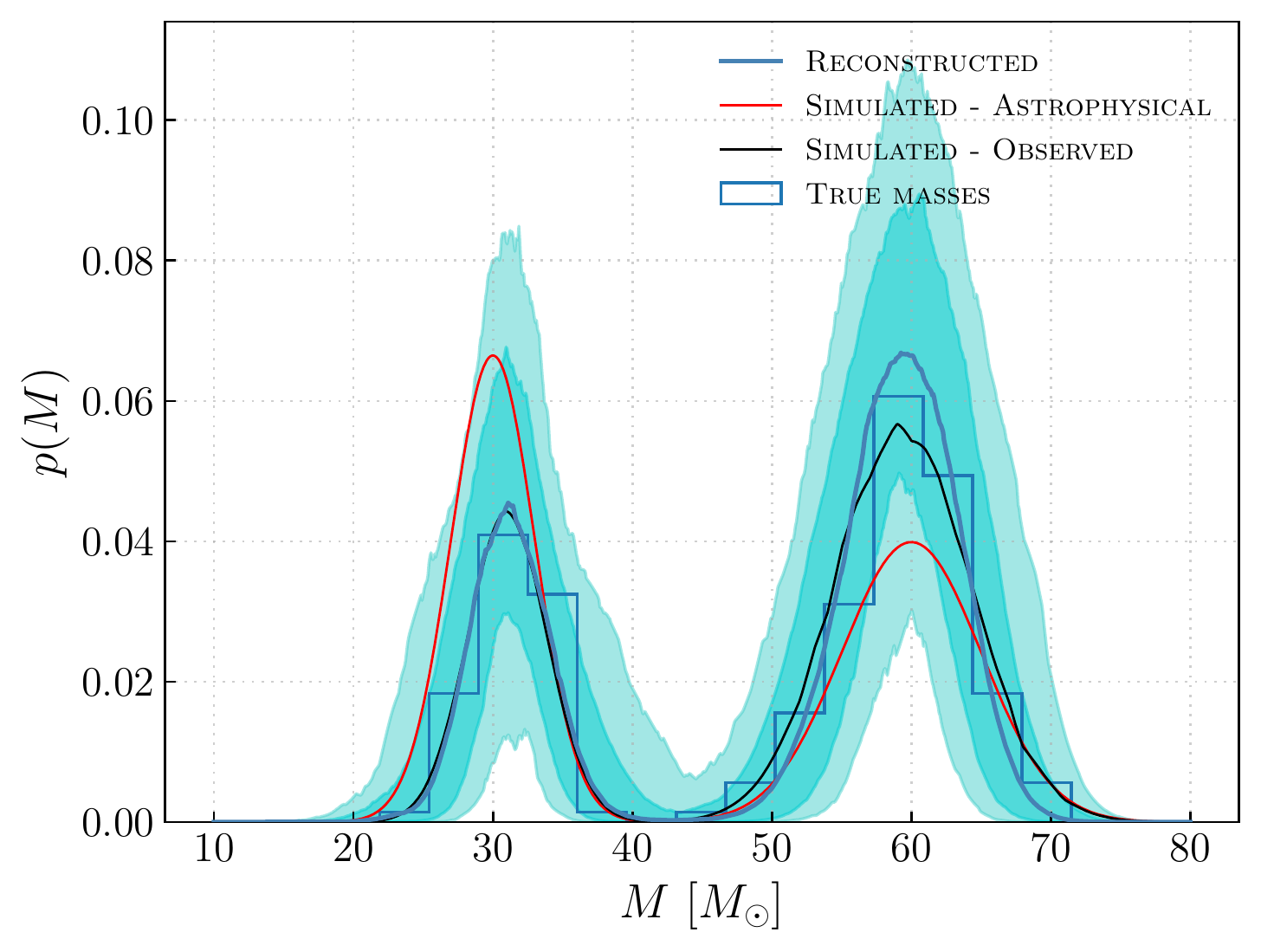}
    \caption{Simulated and recovered observed probability distribution for the bimodal Gaussian case. Corrections to account for selection effects are not applied.}
    \label{fig:seleffects-obs}
\end{figure}

\begin{figure}
    \centering
    \includegraphics[width = \columnwidth]{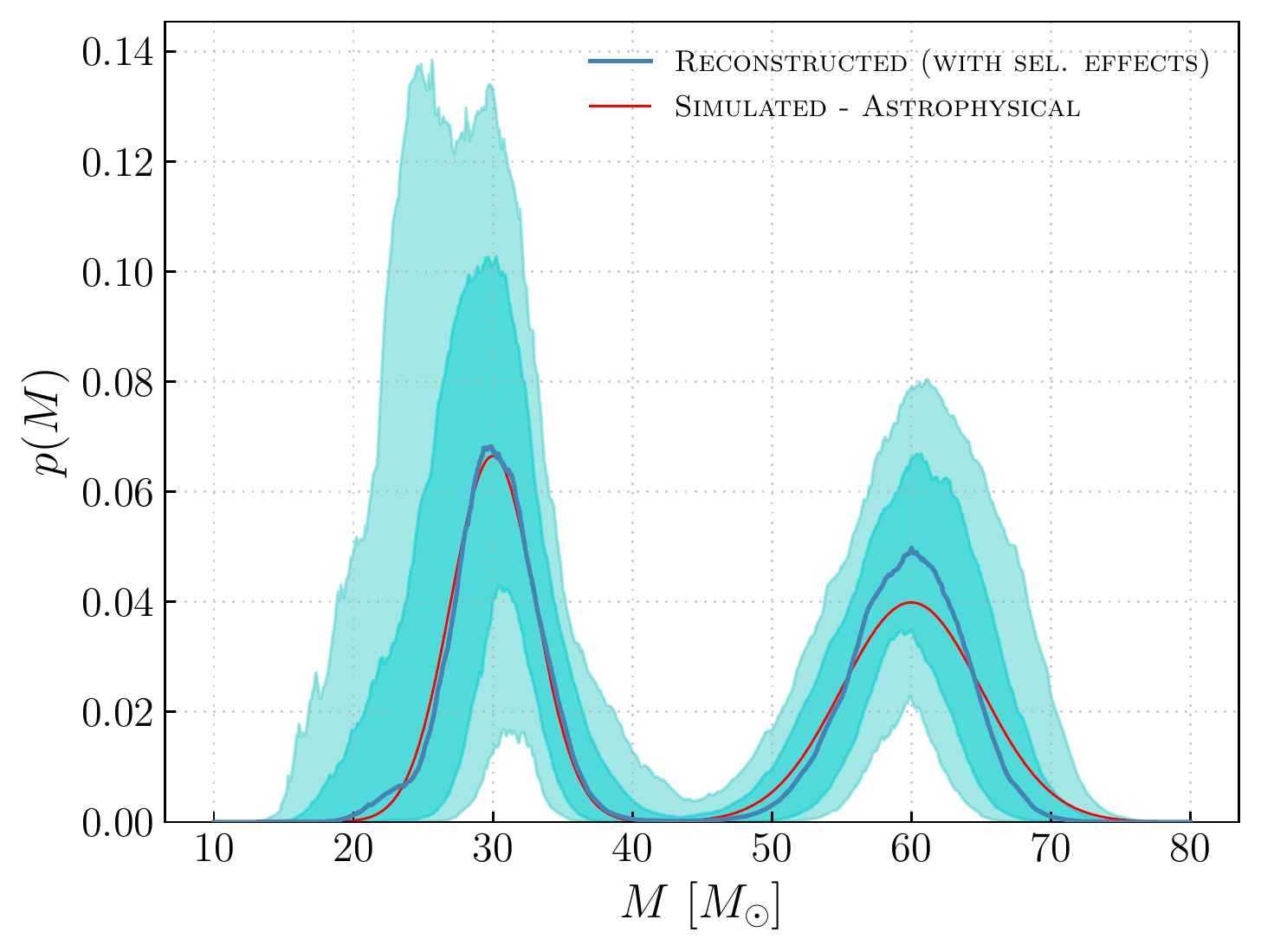}
    \caption{Simulated and recovered astrophysical probability distribution for the bimodal Gaussian case.}
    \label{fig:seleffects}
\end{figure}

The reconstructed observed distribution and the corresponding astrophysical distribution are displayed in Figures~\ref{fig:seleffects-obs} and~\ref{fig:seleffects}. The Jensen-Shannon distance is $\mathrm{JSD} = 0.22^{+0.16}_{-0.11}\ \mathrm{nats}$ from the astrophysical distribution. 

\section{GWTC--2}\label{sec:gwtc}
Having characterised (H)DPGMM on simulated catalogue, we proceed to apply it to the public set of posterior samples from the gravitational wave catalogue GWTC--2~\citep{GWTC2}, available via GWOSC~\citep{gwosc}. We select events following the same prescriptions used in~\citet{pop2}, hence counting only events with false alarm rate (FAR) $< 1\ \mathrm{yr}^{-1}$ and excluding events whose components lie in the NS mass range as well as GW190814~\citep{GW190814}, whose secondary component nature is unknown. 

\begin{figure}
    \centering
    \includegraphics[width = \columnwidth]{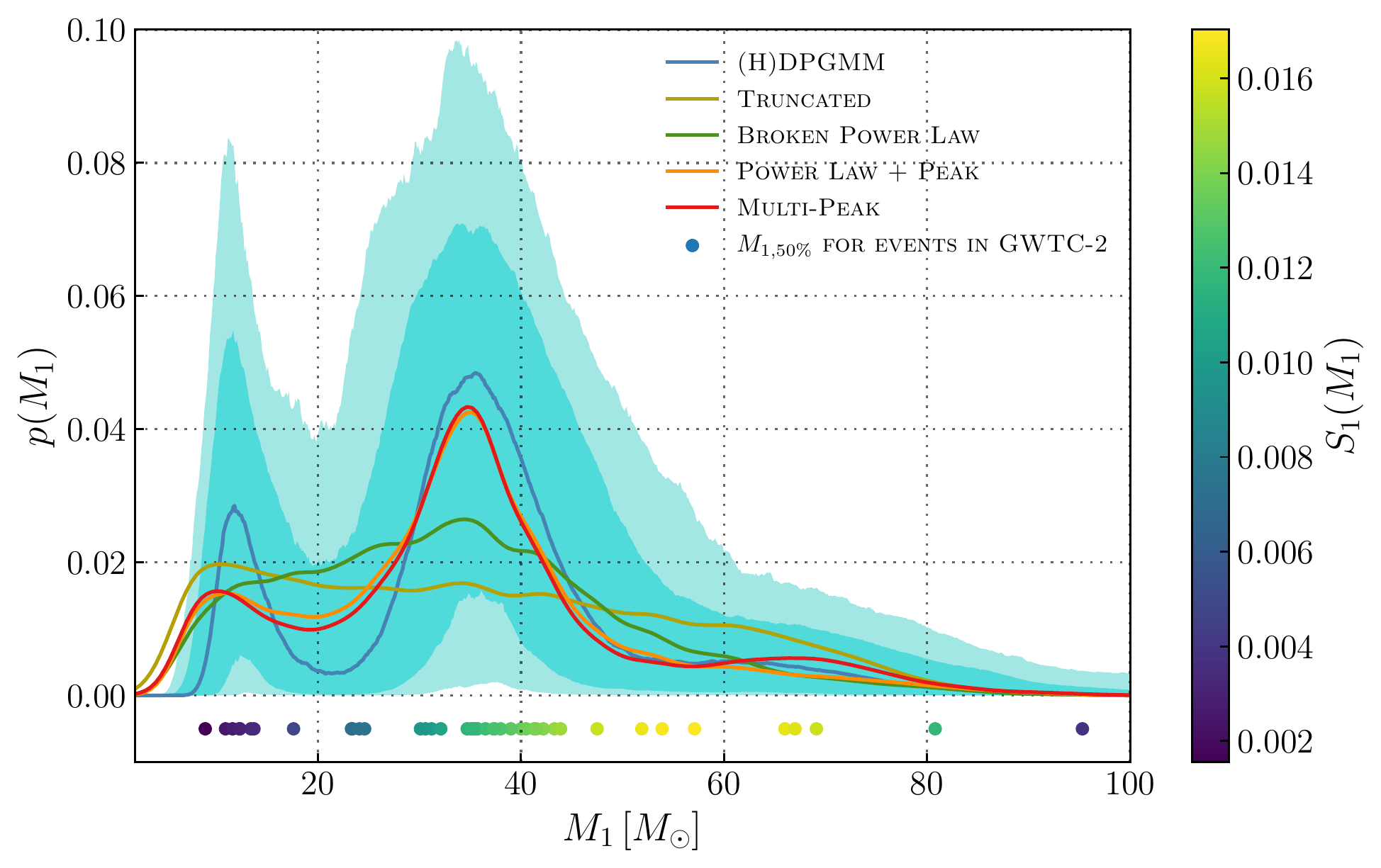}
    \caption{Observed mass function using data from GWTC--2 (primary masses). Coloured bands are $68\%$ and $90\%$ credible intervals for (H)DPGMM. The dots corresponds to median mass values for each event, while the colour gradient is proportional to $S_2(M)$.}
    \label{GWTC-obs}
\end{figure}

\begin{figure}
    \centering
    \includegraphics[width = \columnwidth]{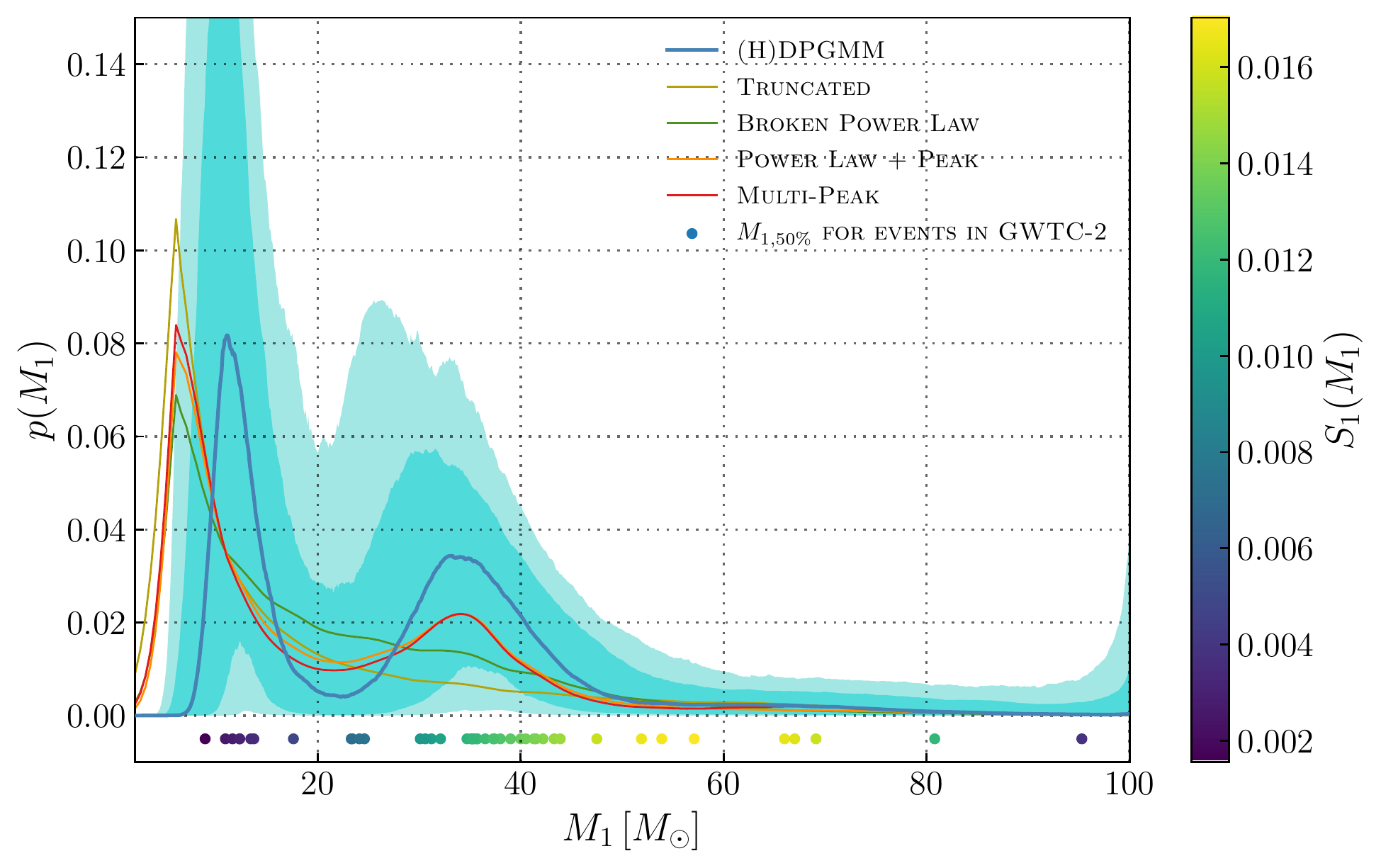}
    \caption{Astrophysical mass function using data from GWTC--2 (primary masses). Coloured bands are $68\%$ and $90\%$ credible intervals for (H)DPGMM. The dots corresponds to median mass values for each event, while the colour gradient is proportional to $S_1(M)$.}
    \label{GWTC-astro}
\end{figure}

To account for selection effects due to the detection thresholds, we used the selection function provided by~\citet{selectionfunction} marginalised over the secondary mass, Figure~\ref{fig:selfunc}. The inferred observed distribution $p(M|O)$ is shown in Figure~\ref{GWTC-obs} where, for ease of comparison, we also report the parametric observed distributions from~\citet{pop2}. Figure~\ref{GWTC-astro} shows instead the inferred astrophysical distribution obtained correcting for selection effects as described above. The parametric models from~\citet{pop2} have also been rescaled according to the same prescription.  
The (H)DPGMM reconstructed mass function is very uncertain, the 90\% credible regions are in fact extremely broad and encompass all parametric models for $M \gtrsim 10 M_\odot$. This is due to the relatively low number of observed events, 43 in total. We do note a significant difference at the low mass end of the inferred mass function. 

Being entirely data driven, (H)DPGMM truncates the mass function at values below the smallest observed mass (GW190924\_021846, $M_1 = 8.9^{+7.0}_{-2.0}\ \msun$). This is stark difference with any parametric models, in which any mass -- within the allowed prior range -- is permitted, regardless of it actually being observed or not. 
As for the features in the observed mass function, we note the presence of two separate main peaks at $\sim 38\ \msun$ and $15\ \msun$ in the (H)DPGMM mass function.

Despite this being indeed suggestive of the presence of two separate populations in merging black holes, when correcting for selection effects these features are smoothed out and we cannot exclude the possibility that they emerge from the limited number of events in the sample. 

To investigate the validity of this hypothesis, we repeated a similar exercise to what presented in Sec.~\ref{sec:undersampling}: we assumed a tapered power law as mass function and generated a total of 45 detected events, repeating the analysis with (H)DPGMM.
We generated the events in two distinct ways:
\begin{enumerate}
    \item sampling 45 individual masses from the underlying mass function;
    \item sampling 45 pairs of masses from the mass function and selecting the largest one from each pair, to mimic the ``$M_1$'' labelling.
\end{enumerate}
In the first (cf. second) case, we find that, out of $10$ realisations, $6\ (7)$ of them do show similar features to the GWTC--2 mass function, although no actual peak was present in the underlying mass function. Hence, we cannot exclude that the apparent presence of two classes of BBH could be due to the limited number of events analysed. 
Note that, however, the relative heights of the two modes of the GWTC--2 distribution differ from our simulations. As the number of detected events increases, we expect these feature to either become statistically significant or to disappear. 

Finally, we note that the (H)DPGMM reconstructed astrophysical mass function seems to grow for $M \gtrsim 100\ \msun$. For such high masses, the selection function goes to zero, therefore no constraints from the data alone are possible and the uncertainty on the mass function itself grows very rapidly.

On a qualitative level, it seems clear the presence of two distinct Gaussian-like populations in the black hole mass function: however, one of the assumptions we made before was that the data themselves are representatives of the underlying distribution. 

In what follows, we will assume that both black holes that composes the merging binaries we observed come from the same formation channels -- there is no difference between them apart from the mass. The fact that we are making a distinction between $M_1$ and $M_2$, considering only the former in our analysis, could mean that we are potentially biasing our analysis,  neglecting some information from the low-mass end of the spectrum.
The primary masses alone, in this picture, are not representative of the underlying distribution, hence our (H)DPGMM reconstruction is not a good approximation of the real mass function.

In order to cure this, we need to include in our analysis the secondary masses as well.
However, due to the fact that the selection functions for primary and secondary mass are different (see Figure~\ref{fig:selfunc}), it is not possible to simply enlarge the sample set with the secondary masses from the observed events.
We decided to analyse the secondary masses separately and, once reconstructed the astrophysical distributions $p_1(M_1)$ and $p_2(M_2)$, join them as
\begin{equation}
    p(M) = \frac{p_1(M)+p_2(M)}{2}\,.
\end{equation}

\begin{figure}
    \centering
    \includegraphics[width = \columnwidth]{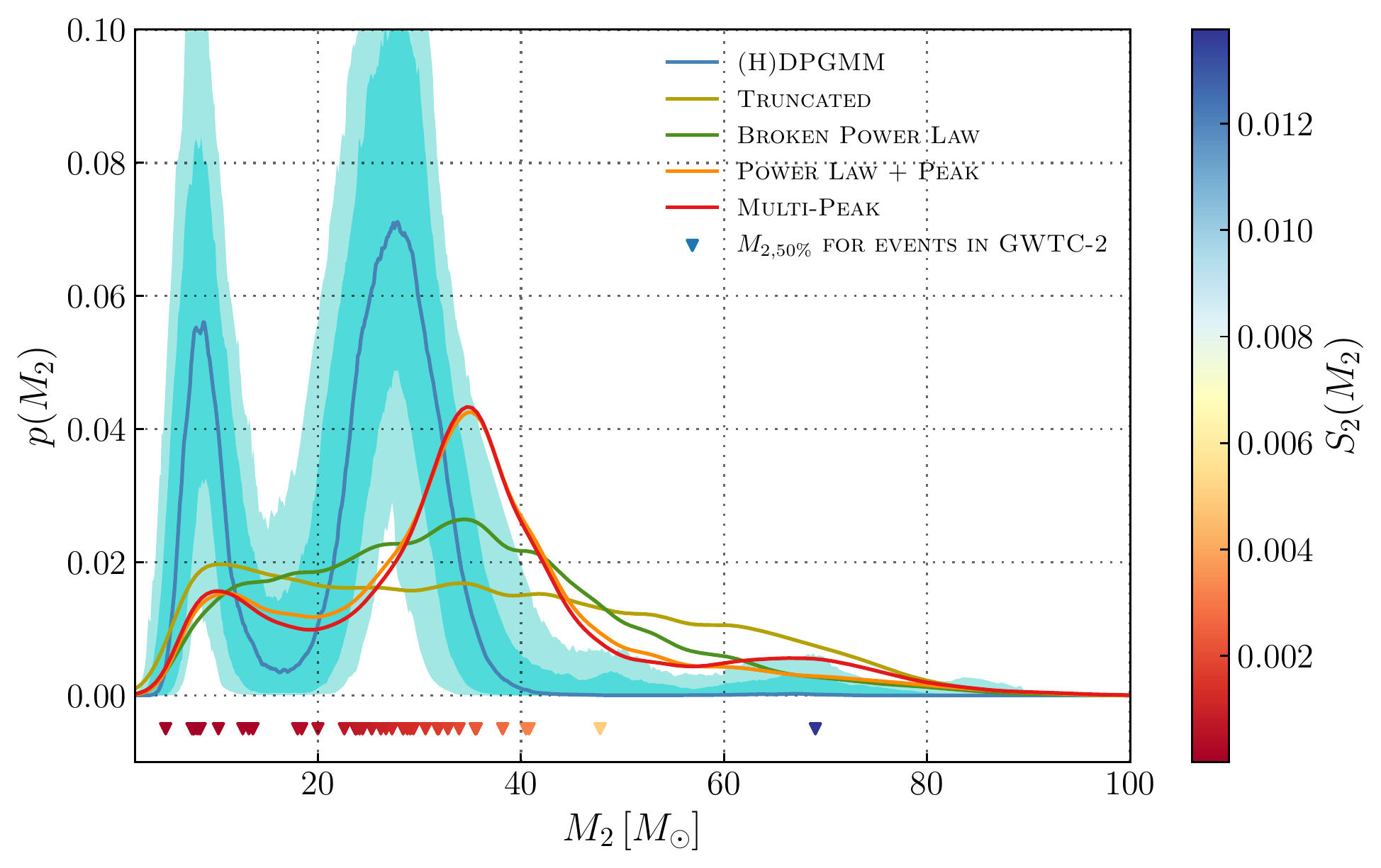}
    \caption{Observed mass function using data from GWTC--2 (secondary masses). Coloured bands are $68\%$ and $90\%$ credible intervals for (H)DPGMM. The triangles corresponds to median mass values for each event, while the colour gradient is proportional to $S_2(M)$.}
    \label{GWTC-m2_obs}
\end{figure}

\begin{figure}
    \centering
    \includegraphics[width = \columnwidth]{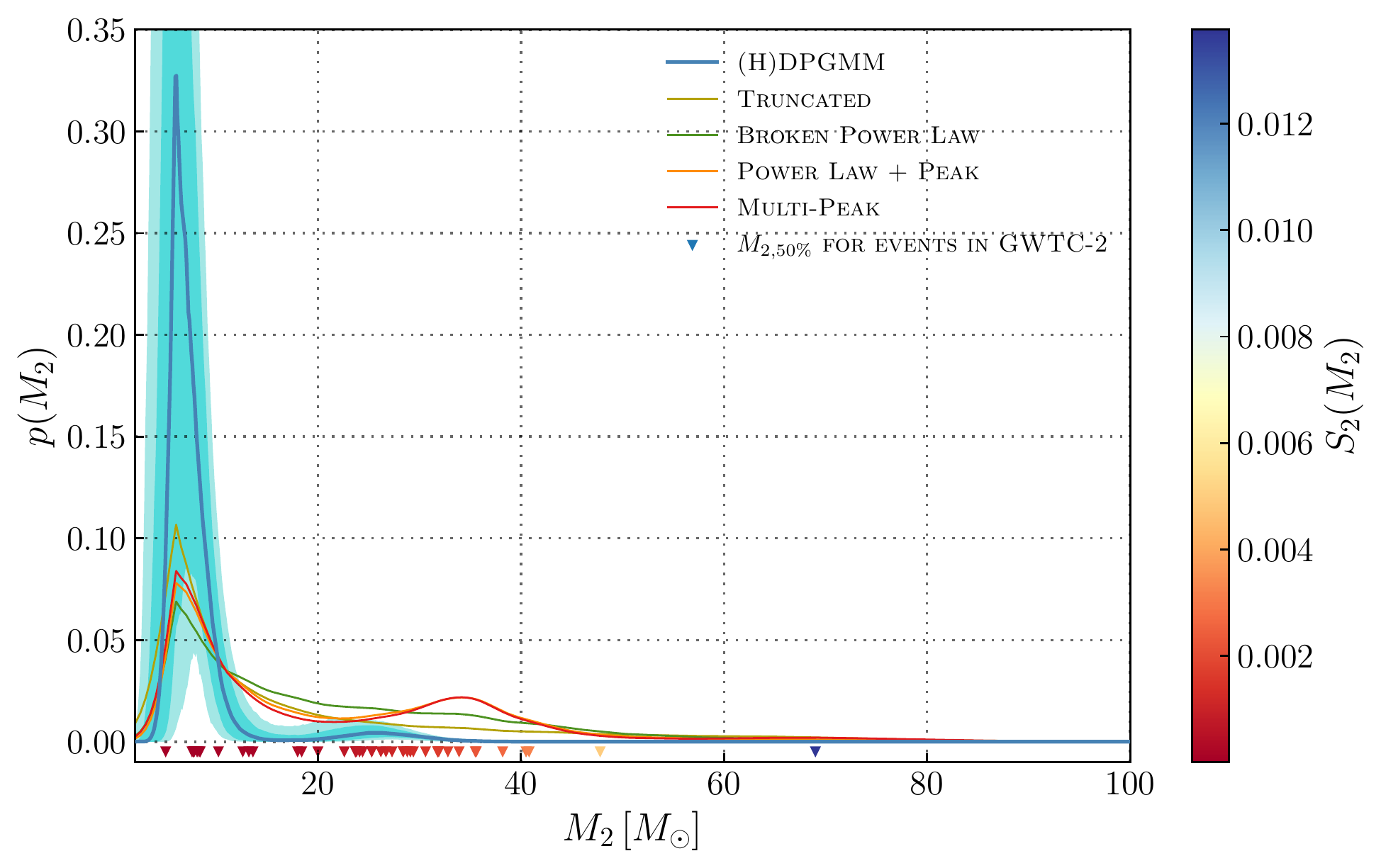}
    \caption{Astrophysical mass function using data from GWTC--2 (secondary masses). Coloured bands are $68\%$ and $90\%$ credible intervals for (H)DPGMM. The triangles corresponds to median mass values for each event, while the colour gradient is proportional to $S_2(M)$.}
    \label{GWTC-m2_astro}
\end{figure}

Figure~\ref{GWTC-m2_obs} shows the observed mass distribution for $M_2$, while Figure~\ref{GWTC-m2_astro} shows the astrophysical distribution for the same quantity. We see that there is, even in this case, a double peak structure in the observed mass function -- once again not compatible with parametric models -- which is almost completely suppressed in the astrophysical distribution.

This is due to the fact that the selection function for the secondary mass goes to zero in the low-mass end of the mass spectrum\footnote{This is reasonable since most of the events with such a low secondary mass will have a low mass ratio, while the events with high mass ratio in this region will have a very low chirp mass: in both cases, the net result is a very low detection probability.}, hence the features in this region are enhanced with respect to the higher-mass features like the secondary peak at $M_2 \sim 25\ \msun$.


\begin{figure*}
    \centering
    \includegraphics[width = 2\columnwidth]{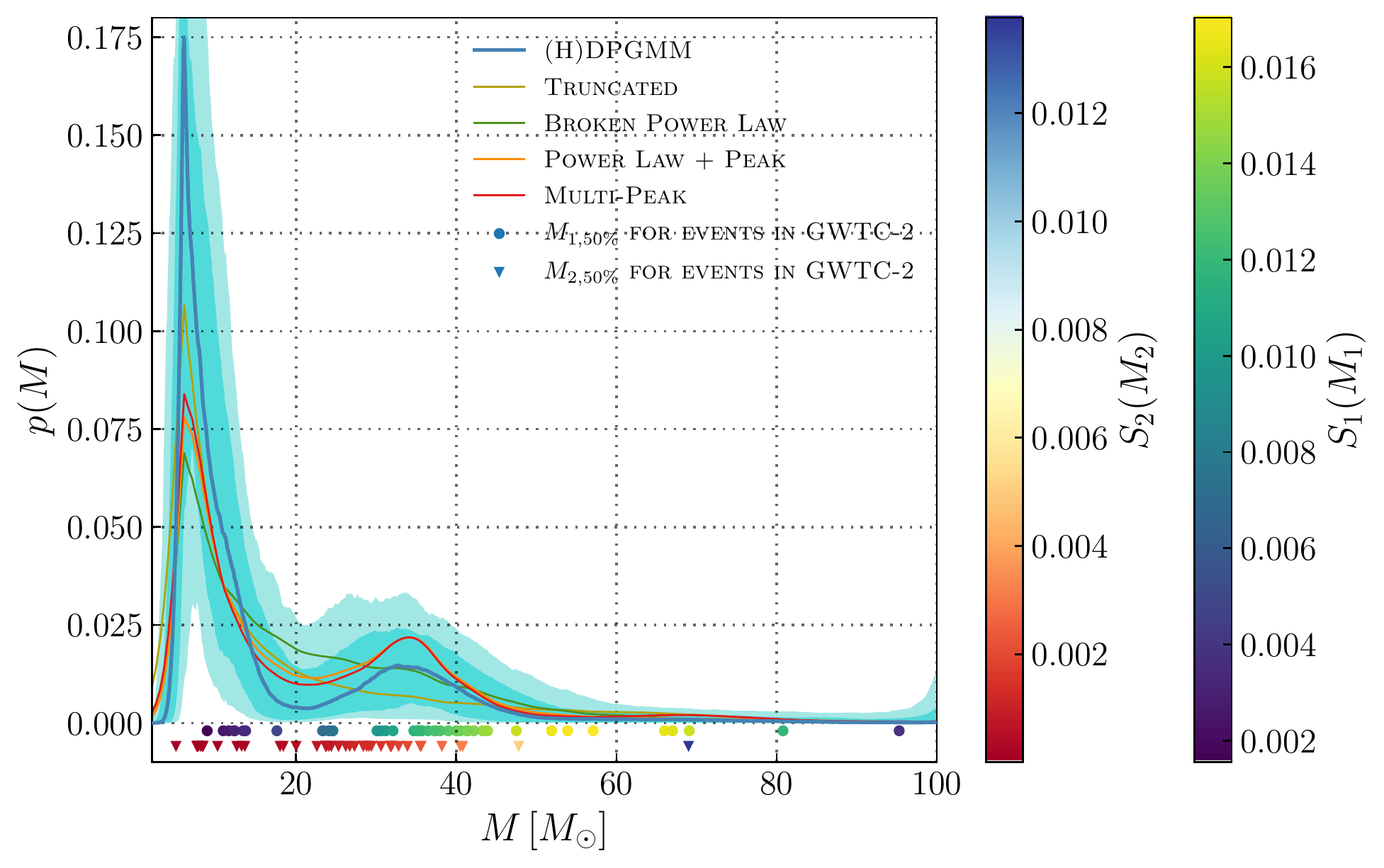}
    \caption{Astrophysical mass function using data from GWTC--2. Coloured bands are $68\%$ and $90\%$ credible intervals for (H)DPGMM. Dots and triangles corresponds to median mass values for each event.}
    \label{GWTC-full}
\end{figure*}

Figure~\ref{GWTC-full} shows the recovered mass function using both primary and secondary masses from GWTC--2. We see that the mass function we reconstructed using (H)DPGMM is compatible with all the parametric models from~\citet{pop2}.

\section{Conclusions}
We presented (H)DPGMM, a non-parametric inference scheme for the merging black hole mass function. Our scheme is based on the DPGMM model, extended to create a hierarchy of non-parametric models. We demonstrated the capabilities of our scheme on controlled simulated events and shown that (H)DPGMM successfully reconstructs a variety of simulated mass functions even when they do not belong to the Gaussian family.

We discussed some of the limitations of our proposed approach; in particular its dependence on the number of observed events as well as the corrections for selection effects. 

We applied (H)DPGMM to GWTC--2 events' primary masses and recovered a mass function different from a tapered power law: in order to discriminate between the possibility that this is due just to the fact that we do not have many low-mass BBH mergers or, on the other hand, that this is a hint towards some new feature in the black hole mass spectrum, more events are required.

The inclusion of the secondary masses in the analysis, under the assumption that both primary and secondary mass come from the same astrophysical distribution, allowed us to infer a probability distribution which is in agreement with the parametric models from~\citet{pop2}.


Data from O3b and from future observing runs (O4 and O5) will help to better understand the shape of the black hole mass function and this non-parametric method could represent a useful guide to build more accurate astrophysically motivated parametric models.

Finally, in this paper, we considered only univariate mixture models. However our algorithm can be easily generalised to allow for the reconstruction of multivariate distributions. Multivariate mixture models will permit investigations in several multidimensional subspaces of the full BBH parameter space to identify correlations and features among parameters other than the mass, e.g. effective spins, that should help further shedding light on the origin of BBH systems. We will investigate such cases in future publications.

\section*{Acknowledgements}
This work benefited from discussion within the Rates \& Population group of the LIGO-Virgo-Kagra collaboration.

This research has made use of data, software and/or web tools obtained from the Gravitational Wave Open Science Center (https://www.gw-openscience.org/), a service of LIGO Laboratory, the LIGO Scientific Collaboration and the Virgo Collaboration. 

LIGO Laboratory and Advanced LIGO are funded by the United States National Science Foundation (NSF) as well as the Science and Technology Facilities Council (STFC) of the United Kingdom, the Max--Planck--Society (MPS), and the State of Niedersachsen/Germany for support of the construction of Advanced LIGO and construction and operation of the GEO600 detector. Additional support for Advanced LIGO was provided by the Australian Research Council. Virgo is funded, through the European Gravitational Observatory (EGO), by the French Centre National de Recherche Scientifique (CNRS), the Italian Istituto Nazionale di Fisica Nucleare (INFN) and the Dutch Nikhef, with contributions by institutions from Belgium, Germany, Greece, Hungary, Ireland, Japan, Monaco, Poland, Portugal, Spain.

\section*{Data availability}
The simulated data underlying this article will be shared on reasonable request to the corresponding author, while GWTC--2 data are available in GWOSC at  \url{https://www.gw-openscience.org/}.

\bibliography{bibliography.bib}

\bsp
\label{lastpage}
\end{document}